\documentclass[12pt,a4paper]{article}

\usepackage[a4paper,margin=1in]{geometry}
\usepackage{newtxtext,newtxmath} 
\usepackage{setspace}
\usepackage{lineno}

\usepackage{amsmath,amssymb}
\usepackage{graphicx}
\usepackage{booktabs}
\usepackage{float}      
\usepackage{array}      
\usepackage{makecell}   
\usepackage{booktabs}
\usepackage{caption}
\usepackage{subcaption}
\usepackage{multirow}
\usepackage{array}
\usepackage{makecell}
\usepackage{float}
\usepackage{url} 
\usepackage{xurl}
\usepackage[hidelinks]{hyperref}
\usepackage{dcolumn}
\usepackage{bm}
\usepackage{enumitem}

\usepackage[T1]{fontenc}
\newcommand{\ditto}{%
  \multicolumn{1}{c}{\textquotedbl}%
}

\usepackage{authblk}

\newcommand{\ws}{\mathrm{WS}}
\newcommand{\fc}{\mathrm{FC}}
\newcommand{\wlp}{\mathrm{WP}}
\usepackage{pdflscape}
\usepackage[round,authoryear]{natbib}

\begin{document}

\begin{titlepage}

\title{\textbf{A systematic evaluation of the Richards equation for predicting soil moisture in Irish grasslands}}

\author[1]{Saurabh Kumar\thanks{Corresponding author: saurabh.kumar@ucd.ie}}
\author[2]{Simon Mathias}
\author[3]{Elodie~Ruelle}
\author[3]{Hussain Khan}
\author[4]{Owen Fenton}
\author[4]{Konstantin Shishkin}
\author[1]{Lennon \'O N\'araigh}
\author[5]{Graham Patrick Benham}

\affil[1]{School of Mathematics and Statistics, University College Dublin, Ireland}
\affil[2]{Department of Engineering, Durham University, United Kingdom}
\affil[3]{Teagasc, Animal \& Grassland Research and Innovation Centre, Moorepark, Fermoy, Cork, Ireland}
\affil[4]{Teagasc, Crops, Environment and Land-Use Programme, Johnstown Castle, Wexford, Ireland}
\affil[5]{Department of Mechanical Engineering, University College London, United Kingdom}
\date{}
\maketitle

\end{titlepage}

\section*{Highlights}

\begin{itemize}
    \item A Richards equation (RE) model was developed for Irish grassland soil moisture prediction.
    \item Hydraulic parameter uncertainty was quantified using Monte Carlo simulations.
    \item A modified Feddes plant stress function was used to account for moisture loss in persisently waterlogged soils.
    \item The RE model was evaluated against observations and an established water-balance model (MoSt).
\end{itemize}

\newpage

\section*{Abstract}

The Richards equation (RE) is widely used to model water flow in unsaturated soils, but its performance in persistently wet grassland systems remains uncertain. This is particularly relevant in Irish grasslands, where soils often remain close to saturation for extended periods and seasonal waterlogging is common.
Here, we evaluate the RE against three soil moisture datasets from County Wexford, Ireland, spanning different locations, soil types, and observation periods. We show that the standard RE formulation systematically over-predicts soil moisture under prolonged near-saturated conditions. We find that this arises from the commonly used Feddes plant water uptake function, which suppresses water losses under anaerobic conditions, despite continued evaporation from near-saturated soils.
To address this limitation, we introduce a simple modification that retains a small non-zero water loss rate in the anaerobic regime. The modified model produces substantially improved agreement with observations across all three datasets.
These results provide a systematic evaluation of RE-based soil moisture modelling in Irish grasslands. More broadly, they identify an important limitation of conventional RE implementations in waterlogged environments and demonstrate a practical approach for improving soil moisture predictions in persistently wet soils.

\vspace{0.5cm}

\noindent\textbf{Keywords:} Richards equation; Grasslands; Grass growth; Soil moisture; Wet soil conditions

\newpage

\section{Introduction}

The Richards equation (RE) is the standard physically based framework for modelling water flow in unsaturated soils. Derived from Darcy's law and conservation of mass, it provides a mechanistic description of infiltration, drainage, evapotranspiration and root water uptake, and has been widely applied across hydrological and agricultural systems~\citep{Simnek2006}. Although originally developed for unsaturated flow, several extensions have been proposed to describe both saturated and unsaturated conditions ~\citep{dogan2005saturated1,dogan2005saturated2}. Despite its widespread use, the predictive performance of the RE depends strongly on the constitutive relationships used to represent soil hydraulic behaviour and plant water uptake~\citep{vanGenuchten}. Consequently, systematic errors may arise when these auxiliary relationships are applied outside the environmental conditions for which they were originally developed and tested.

One such situation occurs in persistently wet soils, as are common in Ireland~\citep{ohara2020}. In such humid environments, prolonged rainfall and shallow water tables can cause large portions of the soil profile to remain close to saturation for extended periods, resulting in seasonal waterlogging. These conditions present a particular challenge for soil moisture modelling because the mechanisms governing water loss differ from those operating in well-drained systems. Relatively little attention has been paid to the performance of RE in such environments where soils remain near saturation for substantial parts of the year.

A key component of the RE is the representation of plant water uptake. This is commonly achieved through a sink term based on the Feddes plant stress function~\citep{FEDDES197613}, in which root water uptake depends on pressure head and vanishes under conditions that are either too dry or too wet for the roots. In standard formulations, water uptake becomes zero once the soil pressure head exceeds a threshold associated with anaerobiosis. This treatment is intended to represent oxygen limitation of plant roots and the consequent suppression of transpiration. However, while transpiration may be strongly reduced under such conditions, evaporation from near-saturated soils can continue. As a result, conventional implementations of the RE may underestimate water losses and overestimate soil moisture when soils remain close to saturation for prolonged periods. Despite the widespread use of Feddes' plant water uptake function, the implications of this assumption for modelling waterlogged soils have received relatively little attention.

Irish grasslands provide an ideal setting in which to investigate this problem. Owing to Ireland's maritime temperate climate, many grasslands experience prolonged periods of near-saturated conditions and seasonal waterlogging. Although Ireland supports an extensive grassland-based agricultural sector~\citep{Dillon2005,Donovan2011,Donovan2022}, no attempt has been made to systematically apply and evaluate the RE using field observations from Irish grasslands. As a result, there remains a limited understanding of how well conventional RE formulations perform under the prolonged wet conditions that frequently occur in these environments.

In Ireland, Teagasc (the National Agriculture and Food Development Authority) has developed a grassland decision support tool called PastureBase Ireland (PBI; \url{https://pasturebase.teagasc.ie}) which has been used by researchers and farmers since 2013~\citep{HANRAHAN2017193,Donovan2022}. In addition, the Moorepark St Gilles Grass Growth (MoSt) model was developed to estimate water availability, which strongly influences predicted grass growth \citep{Ruelle2018, Ruelle2018b,Ruelle_2022}. However, the current soil-water component of MoSt is based on a simplified water-balance approach that does not explicitly resolve vertical soil moisture dynamics (see \citet{BONNARD2025_2,BONNARD2025} for a discussion on the strengths and weaknesses of MoSt). Given the central role of soil moisture in controlling grass production, there is considerable value in evaluating whether physically based approaches such as the Richards equation can provide improved representation of soil-water processes and thereby inform future development of operational grassland forecasting tools.

\subsection{Aim of the paper}

The paper has three aims.   First, we perform a systematic evaluation of the RE for predicting soil moisture dynamics in Irish grasslands using three case studies spanning different locations, soil types and observation periods.  We compare our results with predictions from the MoSt model.   Second, we investigate the causes of discrepancies between model predictions and observations under persistently wet conditions. Third, we propose a modification to the Feddes plant water uptake formulation that retains a small non-zero water loss rate under anaerobic conditions, thereby accounting for continued water losses associated with evaporation from near-saturated soils. We demonstrate that this modification improves agreement with observations for the different case studies.

The work identifies an important limitation of conventional RE implementations in waterlogged environments and provides a simple and practical approach for improving soil moisture predictions in persistently wet soils.  To disseminate use of the model and for reproducibility, the model code is made available via a public GitHub repository.   Monte Carlo simulations are used to account for the uncertainty in the soil hydraulic properties, similarly to \citet{Simnek2006}, \citet{OJHA2014}, and \citet{KUMAR2020125250}.   Improved representation of soil moisture dynamics in persistently wet soils can support better grassland management, with potential benefits for agricultural productivity, environmental sustainability and climate resilience \citep{BOVAL2012,Mayel2021,BONNARD2025}.

\subsection{Layout of the paper}

This paper is laid out as follows. In Section~\ref{sec:MM},  we describe the RE model, and present a summary of the established MoSt model and the inputs to both models -- the soil hydraulic properties and the meteorological data.  We also describe the study sites under investigation as well as the datasets used for the systematic comparisons.   In Section~\ref{sec:results} we present our results, which include a cross-comparison of soil moisture levels based on the observations, the RE model, and the MoSt model.  Discussion and concluding remarks are given in Section~\ref{sec:disc}.

\section{Materials and Methods}
\label{sec:MM}

This section presents the mathematical models, case studies, and methodology used herein. The RE and MoSt models are first described, followed by details of the study sites, datasets, and model evaluation procedures.

\subsection{Mathematical Modeling}
\label{sec:math_model}

In this work, we use two distinct approaches to account for the water balance in the soil column.  The first is the RE model, which is a numerical framework based on the Richards equation, which describes the soil moisture content at all depths, whereas the second is the MoSt model which describes the water storage values aggregated up to $10\,\mathrm{cm}$ and $100\,\mathrm{cm}$ depths only.  We present the mathematical formulation of the RE model and describe its numerical implementation.  Because the MoSt model is well established in the literature, we restrict ourselves here to a high-level summary.

\subsubsection{RE Model}
\label{sec:RE_Model}

\begin{figure}[htb]
\centering

\includegraphics[width=0.5\linewidth]{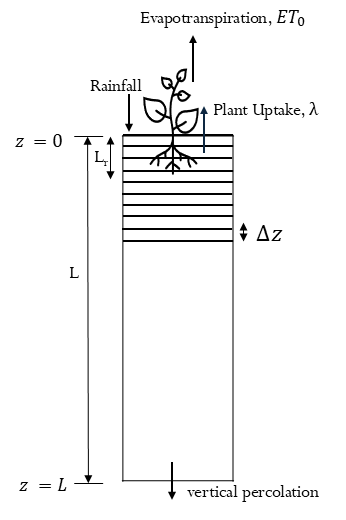}
\caption{Schematic diagram of the Richards Equation (RE) model used to simulate soil water dynamics in Irish grassland systems. Rainfall enters at the soil surface, while root water uptake and evapotranspiration remove water from the root zone ($L_r$). The soil profile, with total depth $L$, is discretized into layers of thickness $\Delta z$.}
\label{fig:schematic}
\end{figure}
For the present analysis, an agricultural field is conceptualized as a 1D vertical model~\citep{Farthing2017} 
as shown in Fig~\ref{fig:schematic}.  The soil profile is represented as a vertical domain from the soil surface to a specified depth.
In the unsaturated zone, flow is governed by the Richards equation~\citep{Richards1931}.  While a 3D numerical model would be more comprehensive, the present study is restricted to 1D simulation due to large data requirements and computational cost associated with 3D models~\citep{Farthing2017, Zha2019}. 

The mixed form of the 1D Richards equation is given as~\citep{Richards1931,celia1990mass},
\begin{equation}
\frac{\partial \theta(h)}{\partial t}
= \frac{\partial}{\partial z} 
\left[K(h)\left( \frac{\partial h}{\partial z} - 1 \right) \right] - \lambda, 
\label{eq:richards_mixed}
\end{equation}
where $\theta(h)$ [-] is the soil moisture, $t$ [$\mathrm{T}$] is time, $z$ [$\mathrm{L}$] is the vertical coordinate (positive downwards), $\lambda$ [$\mathrm{T}^{-1}$] is the sink term associated with plant root water uptake, $K(h)$ [$\mathrm{L}\mathrm{T}^{-1}$] is the  hydraulic conductivity, and $h$ [$\mathrm{L}$] is the pressure head.    Referring to Fig~\ref{fig:schematic}, we adopt the convention whereby the soil profile extends from the lower boundary $z = L$ (bottom of the domain) to the top boundary $z = 0$ (soil surface).

The soil water retention curve, $\theta(h)$ and the hydraulic conductivity curve, $K(h)$ were estimated using the van Genuchten--Mualem relationships~\citep{vanGenuchten} as shown below,
\begin{subequations}
\begin{equation}
S_e=\frac{\theta-\theta_r}{\theta_s-\theta_r} = \left(1+|\alpha h|^n\right)^{-m},\qquad m=1-\frac{1}{n},
\label{eq:VGtheta}
\end{equation}
\begin{equation}
K(h)=K_{sat} S_e^\eta \left[1-\left(1-S_e^{1/m}\right)^m\right]^2,
\label{eq:Kdef}%
\end{equation}%
\end{subequations}%
where $S_e$ [-] is the effective saturation, $\theta_s$ [-] is the saturated soil moisture, $\theta_r$ [-] is the residual soil moisture, $\alpha$ [$\mathrm{L^{-1}}$] is the inverse of air entry pressure head, $K_{sat}$ [$\mathrm{L\,T^{-1}}$] is the saturated hydraulic conductivity, $\eta$ [-] is a pore connectivity parameter and the exponents $n$ and $m$ are dimensionless parameters. Thus, for each soil type, there are six parameters characterizing any given soil: $(\theta_r,\theta_s,\alpha,n, K_{sat},\eta)$. We refer to these six parameters as the van Genuchten (VG) parameters.

We write Equation~\eqref{eq:richards_mixed} in terms of $h(z,t)$ by expanding $\partial \theta/\partial t$ in terms of $\partial h/\partial t$ using a combination of the product rule and the chain rule to account for the compressibility of the soil and water, such that:
%
\begin{equation}
\frac{\partial \theta}{\partial t} =   \left[ (\theta_s - \theta_r) \frac{\partial S_e}{\partial h} + \theta \rho g (c_r + c_w) \right] \frac{\partial h}{\partial t}, 
\label{eq:mixed form}
\end{equation}
where  $\rho$ [$\mathrm{M}\mathrm{L}^{-3}$] is water density, $g$ [$\mathrm{L} \mathrm{T}^{-2}$] is acceleration due to gravity, and $c_r$ [$\mathrm{M}^{-1}\mathrm{L}\mathrm{T}^2$] and $c_w$ [$\mathrm{M}^{-1}\mathrm{L}\mathrm{T}^2$] are compressibilities of soil and water, respectively. The term $\rho g (c_r + c_w)$ is taken as $9.81 \times 10^{-7}\,\mathrm{m}^{-1}$~\citep{Mathias2014}.

In this study, the VG parameters are estimated at different depths based on available soil textural data (sand, silt and clay).  The mean and standard deviation of the VG parameters were obtained using the ROSETTA 3 pedotransfer function which takes soil textural data as inputs~\citep{ZHANG201739}. This was done due to lack of experimentally measured values of these parameters. Pedotransfer functions are commonly used to indirectly determine soil hydraulic properties as they are easy to use and require minimal input data~\citep{Vereecken2015}.  Following this approach, the VG parameters, $\theta_r, \theta_s,$ and $\eta$ were assumed to follow a normal distribution, and $\alpha, n,$ and $K_{sat}$ were assumed to follow a lognormal distribution, each characterized by their mean $(\mu)$  and standard deviation $(\sigma)$.

We describe next the initial and boundary conditions for  Equation~\eqref{eq:richards_mixed}.  
The initial condition  is assumed to be hydrostatic:
\begin{equation}
h(z,t=0) = z - z_{wt}, \qquad 0 \le z \le L, \\
\label{eq:IC}
\end{equation}
where $z_{wt}$ is the depth of the water table measured from the soil surface (usually taken as $z_{wt}=L$ unless otherwise stated). 
At the top boundary, a flux condition is assumed:
\begin{equation}
\begin{aligned}
q(z = 0, t) &= q_i (t), \qquad  t \ge 0,
\end{aligned}
\label{eq:top_flux1}
\end{equation}
where $q=-K(h)(\partial h/\partial z - 1)$ [$\mathrm{L}\mathrm{T}^{-1}$] is the volumetric water flux and $q_i$ [$\mathrm{L}\mathrm{T}^{-1}$] is the infiltration flux on the top surface.  To properly describe the infiltration, ponding logic is used:  if the rainfall rate $q_r(t)$ [$\mathrm{L}\mathrm{T}^{-1}$] is less than the critical ponding flux ($q_p$ [$\mathrm{L}\mathrm{T}^{-1}$]), then the infiltration flux corresponds to the rainfall rate.  Otherwise, once the ponding flux is reached, excess rainfall is rejected, and becomes run-off.  The upper boundary condition is thereby implemented as follows:
\begin{equation}
  q_i = \begin{cases}
    \displaystyle
    q_r,  & q_r \le  q_p, \\
   q_p, & q_r > q_p.
  \end{cases}
  \label{eq:top_flux2}
\end{equation}
The ponding flux  corresponds to the flux which would exist at the top of the soil column in case of saturation $q_p=-K(h)(\partial h/\partial z - 1)_{z=0,S_e=0.999}$, where a near-saturation value of $S_e=0.999$ is used to avoid numerical stability issues.   
In practice, this is done in the context of a finite-difference discretization of the Richards Equation, details of which are given below when the numerical methods are introduced.

At the bottom boundary, depending upon the field conditions, either a free-drainage condition or a zero flow boundary condition is used:
\begin{equation}
  q(z = L, t) =
  \begin{cases}
    \displaystyle
    0,  &\text{zero flow},\\
   K(h), & \text{free drainage}.
  \end{cases}
  \label{eq:vg-k}
\end{equation}

It remains to describe the plant root water uptake term, which constitutes the main sink term, $\lambda$ in Equation~\eqref{eq:richards_mixed}. The plant root water uptake term is defined as~\citep{FEDDES197613}:
\begin{equation}
\lambda=f_1(z)f_2(h)ET_0(t),
\end{equation}
where  $f_1(z)$ [$\mathrm{L}^{-1}$] describes the density of roots, $f_2(h)$ [$-$] is the plant stress function, and $ET_0(t)$ is the potential evapotranspiration [$\mathrm{L} \mathrm{T}^{-1}$].  In this study, daily $ET_0$ values were obtained from Met \'Eireann,  the state meteorological service of Ireland.
The root density is modeled as an exponential function:
\begin{equation}
f_1(z)=\begin{cases}
\frac{a}{L_r}\left[\frac{\exp{(-a)}-\exp({-az/L_r})}{(1+a)\exp({-a})-1}\right],& 0\leq z\leq L_r,\\
  0,& z>L_r.
	\end{cases}
\label{eq:f1def}
\end{equation}
Here, $a$ is a dimensionless parameter and $L_r$ [$\mathrm{L}]$ is the depth below which the roots do not penetrate. The values of $a$ and $L_r$ were assumed as $1\,\mathrm{m}$ and $2\,\mathrm{m}$ respectively~\citep{Mathias2014}. The function, $f_1(z)$ is a true distribution function, in the sense that:
\begin{equation}
\int_0^L f_1(z)\,dz=1,
\end{equation}
which is consistent with the dimensions of $f_1$ being $\mathrm{L}^{-1}$. Furthermore, the plant stress function $f_2(h)$ is modeled as:
\begin{equation}
f_2(h)=\begin{cases}
f_{2a},& h\geq h_a,\\
1,& h_a>h>h_d,\\
1-\frac{h-h_d}{h_w-h_d},& h_d\geq h \geq h_w,\\
0,&h<h_w.
\end{cases}
\end{equation}
The critical pressure heads are $h_a$ for anaerobiosis,  $h_d$ for water-limited evapotranspiration, and $h_w$ for plant wilting point.
The parameter values for this sub-model can be taken from the literature~\citep{Mathias2014, FEDDES197613} and are given here as: $h_a=-0.05\,\mathrm{m}$, $h_d=-4\,\mathrm{m}$, and $h_w=-150\,\mathrm{m}$.

In the standard Feddes model~\citep{FEDDES197613}, the parameter $f_{2a}$ is set to zero, implying that water uptake ceases once the soil reaches the anaerobiosis threshold $h\geq h_a$. This assumption is intended to represent oxygen limitation of plant roots under saturated conditions. However, as discussed earlier, Irish grasslands frequently remain close to saturation for extended periods of the year, and evaporation from near-saturated soils may continue even when transpiration is strongly suppressed. Under such conditions, complete suppression of plant water uptake can lead to systematic over-prediction of soil moisture by the Richards equation. To investigate this hypothesis, we modify the Feddes function by assigning a small non-zero value to $f_{2a}$. This modification provides a simple representation of residual water losses under near-saturated conditions while preserving the simplicity of the original formulation. In the present study, $f_{2a}=0.01$ is adopted as the baseline value. The sensitivity of model predictions to $f_{2a}$, and the extent to which this modification improves agreement with observations, are examined in Appendix~A.


\subsubsection{Numerical implementation}

The governing equations are discretized in space using a finite-difference scheme on a uniform grid of size $\Delta z$,  with $z_i=i\Delta z$ and $i\in \{1,\cdots, N\}$, where $N$ is the number of nodes in the discretization.  Correspondingly, we denote $h_i=h(z_i,t)$.  This reduces the Richards equation to a set of ordinary differential equations (ODEs), $d h_i/d t=F_i(h_1,\cdots h_N)$.  These ODEs are solved in Python using the \texttt{solve\_ivp} function from the  \texttt{scipy.integrate} Python library.   The \texttt{bdf} algorithm is chosen, as this is based on implicit backward differentiation with adaptive timestepping.
We have used a grid spacing $\Delta z = 0.02\,\mathrm{m}$ over a soil column of depth $L = 2\,\mathrm{m}$. A grid sensitivity analysis was performed by varying $\Delta z$, and it was found that $\Delta z = 0.02\,\mathrm{m}$ is sufficient for a grid-independent solution.  

In the context of the finite-difference method, we define the ponding flux in precise quantitative terms (the ponding flux arises in the context of the top boundary condition at $z=0$, as per Equation~\eqref{eq:top_flux2}).  From the VG function, we compute $h_*$ as the value of the pressure head corresponding to $S_e=0.999$ (the use of $h=0$ at saturation  is avoided here, to promote the stability of the numerical method~\citep{Mathias2014}).  Then, the ponding flux is computed as:
\begin{equation}
q_p=-K(h_*)\left(\frac{h_*-h_1}{\Delta z}-1\right),
\label{eq:qpdef}
\end{equation}
where $h_1$ is the value of $h$ at the first finite-difference node just below the surface ($z=0$).

As a first validation step, our RE implementation was applied to the different case studies for different soil types reported by \citet{Mathias2014} and excellent agreement was obtained, giving confidence in the present numerical implementation. Example code from our RE model implementation is available in a Github repository, details of which are given later in the Data Availability Statement.

To account for uncertainty in the VG parameters, we use Monte Carlo simulations.  Hence, for each case study involving a time series of $\theta(\cdot,t)$, we present not only a mean trajectory, but also a confidence interval.  This is achieved as follows.  For each case study involving a given soil type, the mean and the standard deviation of the VG parameters as determined by ROSETTA 3 is obtained.  We then draw a set of VG parameters at random from the distribution characterized by this mean and standard deviation, and carry out a simulation of the RE model, to produce a particular realisation of $\theta(\cdot,t)$.   We repeat this procedure, drawing different VG parameters at random each time, to build up an ensemble of simulations, from which the mean trajectory and the standard deviation can be obtained.  
Using this approach,  a convergence analysis was performed to determine the minimum number of realizations needed to produce converged mean values of $\theta$ and $h$; this was found to be 200. 

\subsubsection{Derived Quantities}

The water storage $(\Theta)$ in the soil column can be calculated from $\theta$ in a standard fashion:
\begin{equation}
\Theta(t) =\int_0^L \theta(z,t) \,dz.
\label{eq:water_storage}
\end{equation}
The water storage can be estimated from  the finite-difference discretization of $\theta$ via the trapezoidal integration rule as:
\begin{equation}
\Theta(t) =\left( \frac{\theta(z_1,t)+\theta(z_N,t)}{2} + \sum_{i=2}^{N-1} \theta(z_i,t)\, \right)\Delta z.
\label{eq:waterstorageREmodel}
\end{equation}
Here, the sum ranges over all gridpoints in the finite-difference discretization of the computational domain, $z\in [0,L]$ (with the first and last terms halved).  Correspondingly, the water storage from $z=0$ to a depth $\ell<L$ is written as $\Theta_\ell(t)$ and can be obtained from Equation~\eqref{eq:waterstorageREmodel} by summing only over those gridpoints $z_i\in [0,\ell]$.
A similar approach is used when estimating $\Theta(t)$ and $\Theta_\ell(t)$ from the observations, which are provided at various depths.


For the purpose of comparing with observations, it will be useful to convert between water tension values $\psi\,(\mathrm{hPa})$ and pressure-head values $h\,(\mathrm{m})$.  The conversion is given by:
\begin{equation}
\psi = -\rho g h.
\label{eq:waterTensionPressureHead}
\end{equation}
Thus,  $\psi > 0$ in an unsaturated soil  and $\psi<0$ in a saturated soil.

\subsubsection{MoSt Model}

The MoSt model is a dynamic, mechanistic model developed by Teagasc to help farmers with efficient grassland use~\citep{Ruelle2018,Donovan2022}. It predicts daily grass growth, plant nitrogen content and flux as a function of soil type, fertilization and pasture management~\citep{Ruelle2018}. It was developed in C++ as a grassland decision support tool~\citep{Ruelle2018}. 

The model is composed of several interconnected sub-models representing grass growth, soil water and nitrogen fluxes, and the animal presence on the paddock. The estimation of  grass growth is based on the calculation of potential growth depending on solar radiation and the grass biomass that is then scaled using parameters depending on environmental conditions (temperature, water availability, nitrogen availability, and radiation). 

The initial MoSt model was based on an earlier grass growth model~\citep{Jouven2006}. Later, the soil and water components were refined to better represent different soil types and to account for different soil depths, improving the simulations of water and soil nitrogen fluxes~\citep{BONNARD2025_2}. \citet{BONNARD2025_2} also added a topsoil sub-layer which improved the model accuracy in drier years. Further details of the MoSt model are described by~\citet{Ruelle2018} and~\citet{BONNARD2025_2}. In the present study, we focus on the soil and water sub-model in the MoSt model. 

The main inputs to the model include soil texture (sand and clay), organic matter content, depth, daily meteorological data (rainfall $(q_r)$, solar radiation, temperature, and $ET_0$), nitrogen fertilization, and grazing/cutting management. The model outputs include daily grass growth (kg DM ha$^{-1}$ day$^{-1}$), biomass (kg DM ha$^{-1}$), grass height (cm), grass nitrogen content (g N kg$^{-1}$ DM), nitrate leaching (kg N ha$^{-1}$), soil mineral nitrogen and organic nitrogen content (kg N ha$^{-1}$) and soil water storage. These outputs can be obtained on a daily, weekly, monthly or yearly basis, or the average of the yearly simulation~\citep{Ruelle2018}.


\subsection{Datasets and Case Studies}

\begin{figure}[htb]
	\centering
		\includegraphics[width=0.7\textwidth]{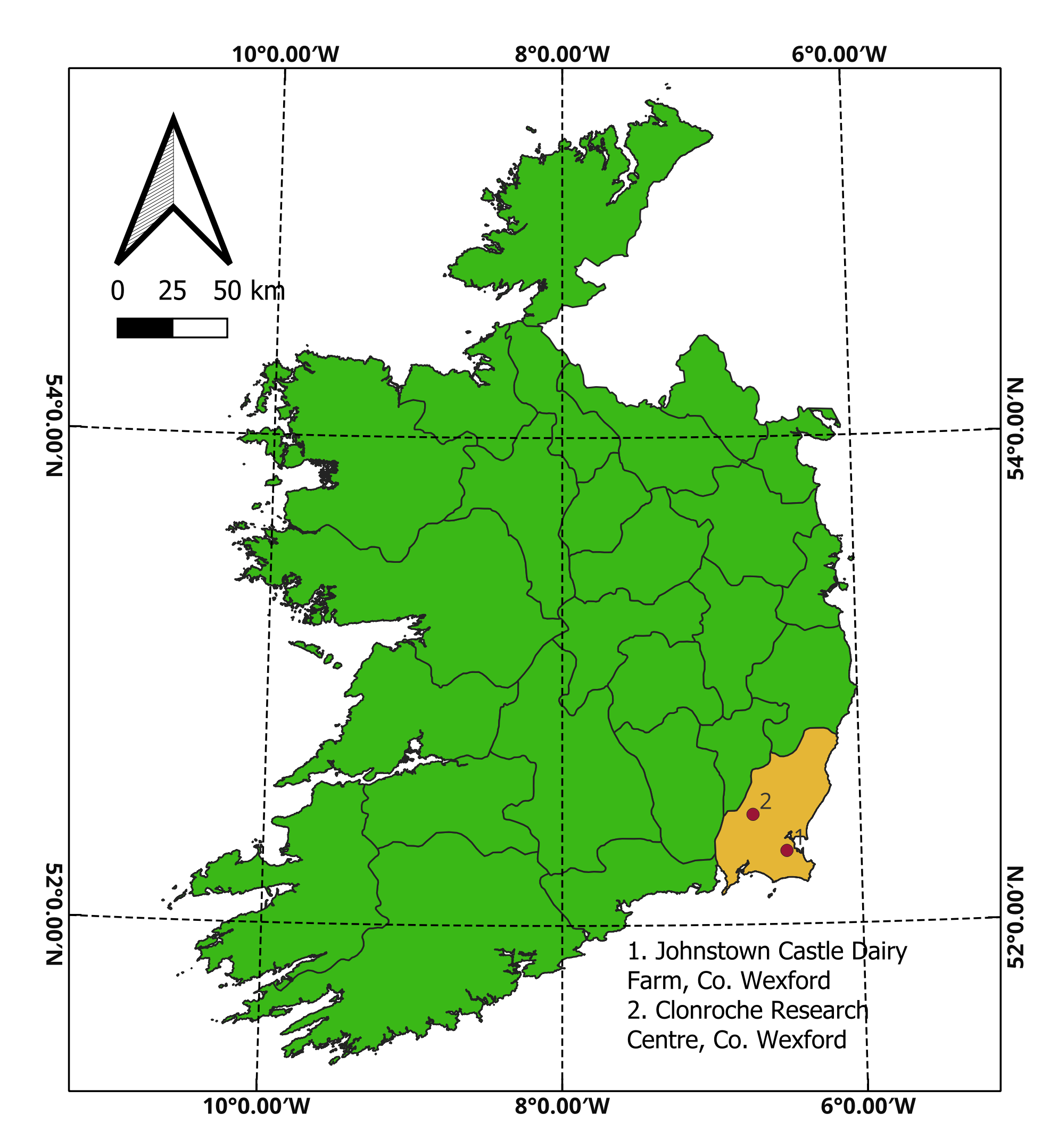}
		\caption{Map of Ireland showing the location of the study sites in County Wexford, Ireland}
	\label{fig:map}
\end{figure}

The present work is based on the two datasets outlined in Table~\ref{tab:study}.  The datasets involve two study sites and three distinct case studies.  The study sites are shown in  Fig~\ref{fig:map}.  The case studies correspond to soils with high, medium, and low clay content,  which we quantify below.

\begin{table}[htb]
\centering
\small 
\caption{Summary of case studies. D\&S refers to the dataset by \citet{Diamond2001}. ISMON refers to the dataset from the Irish Soil Monitoring Network~\citep{ISMON2021}.}
\begin{tabular}{ccccccc}
\hline
\hline
Reference & Dataset & Study Site  & Clay Content & Measurement & Variable & Case Study \\ 
\hline
D\&S  & 1 & Clonroche           & medium & Tensiometer &  $\psi$ & 1 \\
D\&S  & 1 & Johnstown Castle    & low    & Tensiometer & $\psi$ & 2 \\
ISMON & 2 & Johnstown Castle    & high   & TDR         & $\theta$ & 3 \\
\hline
\hline
\end{tabular}
\label{tab:study}
\end{table}

\paragraph*{Diamond and Sills Dataset (including meteorological data):}

Between 1998 and 2000, \citet{Diamond2001} measured daily soil water tension at three different sites. The measurements were subsequently build a soil-moisture deficit model for Ireland~\citep{schulte2005predicting}.  The sites were: Ballintemple Nursery (County Carlow), Clonroche Research Station (County Wexford), and Johnstown Castle Research Centre (County Wexford). Pressure transducers and tensiometers were installed at each site to measure soil water tension $(\psi)$ values at depths of $15\,\mathrm{cm}$, $30\,\mathrm{cm}$, $45\,\mathrm{cm}$, $60\,\mathrm{cm}$, $90\,\mathrm{cm}$,  and $120\,\mathrm{cm}$.   These values are presented by \citet{Diamond2001} and are extracted using a plot digitizer.   Clonroche and Johnstown Castle each host a Met \'Eireann weather station, meaning that a quantitative description of evapotranspiration is available at these sites to support the RE model.  Ballintemple Nursery does not host such a weather station and for that reason is excluded from the present study.

At Clonroche, pressure transducers with an operating range of $0$ to $-1$ bar were used. The transducers had an operating temperature range of $0^\circ$ to $60^\circ$ C, a linearity of $25\%$ full scale and a hysteresis of less than $1\%$.
At Johnstown Castle, pressure transducers with a stated compensated temperature range of $-20^\circ$ to $80^\circ$C were used. The tensiometers were refilled with water twice weekly. To account for the water column in the tensiometer tube, measurements were corrected by subtracting the corresponding pressure head. This correction can result in negative (saturated) values, indicating the presence of a water table within the soil profile~\citep{Richards1965}.

\citet{Diamond2001} also measured the soil texture for Clonroche and Johnstown Castle.  Based on these measurements, we use  ROSETTA 3 to estimate the mean ($\mu$) and standard deviation ($\sigma$)  of the VG parameters for Clonroche and Johnstown Castle.  Results of this analysis are presented in Tables~\ref{tab:soil_data_CR1}--\ref{tab:soil_data_JC2} below (all tables are grouped together at the end of this section, for readability).
Using the values from these tables, together with observed water-tension values, we have used the VG model to estimate $\theta$ at different depths. 
Using the daily values of soil moisture and water tension, we have also estimated the depth-averaged soil moisture $(\bar\theta)$ and water tension $(\bar \psi)$.  
These daily depth-averaged values are presented for all case studies in Section~\ref{sec:results}.


At Clonroche and Johnstown Castle, there is a weather station operated by Met \'Eireann.  Historical daily values of $q_r$ and $ET_0$ are available for these sites on the Met \'Eireann website (\url{https://www.met.ie/climate/available-data/historical-data}).  We use these values as inputs into the RE model. Met \'Eireann calculates $ET_0$ using the FAO-56 Penman--Monteith equation. 

For the Diamond and Sills dataset, $ET_0$ data were not available between 1998 and 2000, while the rainfall data were available. As per the recommendation of FAO-56, the Hargreaves equation was used to estimate potential evapotranspiration from temperature data and a calibration relationship was derived between the evapotranspiration values estimated using the Hargreaves and Penman--Monteith equations~\citep{allen1998fao}. The relationship is given as follows:
\begin{equation}
ET_{\text{Penman--Monteith}} = 0.375\, ET_{\text{Hargreaves}}.
\label{eq:CalibrationEq}
\end{equation}
The $R^2$ and RMSE for the calibration relationship were $0.85$ and $0.410$ respectively.   
In this way, the above calibration was used to estimate $ET_0$ for the period 1998-2000.

\paragraph*{ISMON Dataset (including meteorological data):}  The Irish Soil Moisture Monitoring Network (ISMON) is a long-term soil moisture observation platform launched in 2021~\citep{ISMON2021}. ISMON collects hourly soil moisture data across 10 sites in Ireland representing a range of soil types and land uses. At each ISMON station, barometric pressure, relative humidity, air temperature, wind velocity, rainfall, and net solar radiation are measured. 

The network consists of multiple  time-domain reflectometry (TDR) sensors installed in soil pits excavated around the Cosmic Ray Neutron Sensors (CRNS) stations. After installation, the soil pits were refilled. The TDR sensors were arranged in radial arrays extending from the CRNS stations at azimuths of $0^\circ$, $120^\circ$, and $240^\circ$, with installation distances of approximately $20\,\mathrm{m}$ from the center of the sensor footprint. 
Within each array, TDR probes were installed horizontally at depths of $5\,\mathrm{cm}$, $10\,\mathrm{cm}$, $20\,\mathrm{cm}$, $30\,\mathrm{cm}$, $50\,\mathrm{cm}$, and $100\,\mathrm{cm}$, covering the upper soil profile and extending below the drainage layer where feasible. We obtained the hourly soil moisture $(\theta)$ of Johnstown Castle from ISMON and the daily meteorological  from Met \'Eireann.   This is the same research farm as in the earlier study by~\citet{Diamond2001}, but the soil types are different.  The Diamond and Sills dataset involves a soil with a low clay content, whereas the ISMON dataset involves a soil with a high clay content.

The soil at the site was classified as imperfectly drained, consisting of a fine loamy topsoil over stony subsoils with increasing rock fragment content with depth. Agricultural gravel drains are installed at depths of approximately $30-50\,\mathrm{cm}$. The parent material comprises quartzite tills over Cambrian greywacke, slate, and quartzite bedrock, with Stagnic Luvisols being the dominant soil type. The soil textural data were collected up to $30\,\mathrm{cm}$ depth and are given in Table~\ref{tab:soil_data_ISMON1}. The soil texture data were obtained using Laser Diffraction Particle Size Analysis.
Using ROSETTA 3, the mean ($\mu$) and standard deviation ($\sigma$)  values of the VG parameters were  again estimated.  The values are given below in Table~\ref{tab:soil_data_ISMON2}.  Values in brackets are adjusted values, to account for the presence of gravel drains at depths between $0.3\,\mathrm{m}$ and $0.5\,\mathrm{m}$.
Using the inverse VG relationship, the $h$-values were estimated at different depths. Using the hourly $\theta$ and $h$ values, the depth-averaged soil moisture $(\bar\theta)$ and pressure head $(\bar h)$ values were estimated. 
Hourly values of $\bar\theta$ and $\bar\psi$ are presented in Fig~\ref{fig:MetISMON}.   We also show the corresponding meteorological data (daily rainfall and $ET_0$) in this figure.

\paragraph*{Distinction between the datasets:}  We emphasize that the earlier dataset of Diamond and Sills contains observations of pressure head $h$ at various depths, on an hourly basis.  From these observations, we have inferred values of $\theta$, using the VG model.  In contrast, in the ISMON dataset, hourly values of soil moisture $\theta$ at various depths are provided, from which we infer values of $h$, again using the VG model.  In this way, in the two datasets, the `raw' observations are taken with respect to different physical quantities.  For this reason, in the following sections, when comparing our model with the observations, we treat the two datasets separately.

\subsection{Methodology}
The present study aims to estimate the water storage, $\Theta$, from 0--10 cm depth $(\Theta _{10})$ and from 0--100 cm depth $(\Theta _{100})$ using both the RE and MoSt models. The methodology is outlined below:

\begin{itemize}[noitemsep]
    \item Daily rainfall $q_r$ and $ET_0$ for different study areas were obtained from Met \'Eireann. 
    \item Soil textural data (sand (\%), silt (\%), and clay (\%)) were used to estimate the mean and standard deviation of VG parameters ($\theta_s$, $\theta_r$, $\alpha$, $n$, $\eta$, $K_{sat}$) generated using ROSETTA 3~\citep{ZHANG201739} at different depths. It was assumed that $\theta_s$, $\theta_r$, and $\eta$ follow normal distributions while $\alpha$, $n$ and $K_{sat}$ follow lognormal distributions. The VG parameters for the different sites are shown in Tables~\ref{tab:soil_data_CR2}, \ref{tab:soil_data_JC2}, and~\ref{tab:soil_data_ISMON2}.
    \item The RE model was run with VG parameters, daily rainfall $q_r$ and $ET_0$ as input, and daily pressure head and soil moisture were obtained as outputs. The water storage values up to depths of 10 cm $(\Theta_{10})$ and 100 cm $(\Theta_{100})$ were computed using Equations~\eqref{eq:water_storage}-\eqref{eq:waterstorageREmodel}. For depths beyond the available soil data, the VG parameters of the deepest layer were extrapolated downwards assuming the same values. 
    \item The MoSt model was run with soil textural data, rainfall, and $ET_0$ as inputs and the $\Theta _{10}$ and $\Theta _{100}$ values were obtained as outputs.
    \item From the observed datasets, the $\Theta _{10}$ and $\Theta _{100}$ values were computed using trapezoidal integration method as per Equations~\eqref{eq:water_storage}-\eqref{eq:waterstorageREmodel}. 
    \item After computing water storage values from the RE and MoSt models, they were compared against the observed datasets.  

\end{itemize}

\newpage

\begin{landscape}

\begin{table}[htb]
\centering
\caption{Soil texture at different depths, Clonroche -- \citet{Diamond2001} dataset}
\label{tab:soil_data_CR1}
\begin{tabular}{lcccccc}
\hline
\hline
Site & Depth min (m) & Depth max (m) & Soil type & Sand (\%) & Silt (\%) & Clay (\%) \\
\hline
\hline
\multirow{4}{*}{Clonroche} & 0.00 & 0.25 & clay loam & 29 & 39 & 32 \\
 & 0.25 & 0.60 & clay loam & 31 & 40 & 29 \\
 & 0.60 & 0.90 & loam & 36 & 38 & 26 \\
 & 0.90 & 1.20 & clay loam & 29 & 42 & 29 \\
\hline
\hline
\end{tabular}
\end{table}

\begin{table}[H]
\centering
\small

\caption{Mean ($\mu$) and standard deviation ($\sigma$) of VG parameters at different depths for the \citet{Diamond2001} dataset at Clonroche. The units of $\alpha$ and $K_{sat}$ are m$^{-1}$ and m/day, respectively. All other VG parameters are dimensionless.}
\label{tab:soil_data_CR2}

\begin{tabular}{cccccccccccccc}
\hline
\hline

\multicolumn{2}{c}{Depth (m)} &
\multicolumn{2}{c}{$\theta_s$} &
\multicolumn{2}{c}{$\theta_r$} &
\multicolumn{2}{c}{$\log_{10}(\alpha)$} &
\multicolumn{2}{c}{$\log_{10}(n)$} &
\multicolumn{2}{c}{$\log_{10} (K_{sat})$}  & 
\multicolumn{2}{c}{$\eta$}  \\

Min  & Max & $\mu$ & $\sigma$ & $\mu$ & $\sigma$ & $\mu$ & $\sigma$ & $\mu$ & $\sigma$ & $\mu$ & $\sigma$ & $\mu$ & $\sigma$ \\
\hline
\hline
0.00 & 0.25 & 0.427 & 0.008 & 0.110 & 0.008 & -2.190 & 0.075 & 0.137 & 0.011 & 0.967 & 0.123 & -0.506 & 1.548\\
0.25 & 0.60 & 0.421 & 0.008 & 0.106 & 0.007 & -2.207 & 0.074 & 0.142 & 0.012 & 0.982 & 0.126 & -0.434 & 1.500\\
0.60 & 0.90 & 0.412 & 0.008 & 0.101 & 0.007 & -2.166 & 0.077 & 0.142 & 0.013 & 0.964 & 0.124 & -0.490 & 1.394\\
0.90 & 1.20 & 0.425 & 0.009 & 0.106 & 0.008 & -2.238 & 0.077 & 0.145 & 0.012 & 1.004 & 0.128 & -0.359 & 1.535\\
\hline
\hline
\end{tabular}

\end{table}

\end{landscape}

\newpage

\begin{landscape}

\begin{table}[htb]
\centering
\caption{Soil texture at different depths, Johnstown Castle -- \citet{Diamond2001} dataset}
\label{tab:soil_data_JC1}
\begin{tabular}{lcccccc}
\hline
\hline
Site & Depth min (m) & Depth max (m) & Soil type & Sand (\%) & Silt (\%) & Clay (\%) \\
\hline
\hline
\multirow{4}{*}{Johnstown Castle} & 0.00 & 0.25 & loam & 49 & 32 & 19 \\
 & 0.25 & 0.60 & loam & 49 & 33 & 18 \\
 & 0.60 & 0.90 & loam & 36 & 47 & 17 \\
 & 0.90 & 1.20 & loam & 51 & 36 & 13 \\
\hline
\hline
\end{tabular}
\end{table}

\begin{table}[H]
\small
\centering
\caption{Mean ($\mu$) and standard deviation ($\sigma$) of VG parameters at different depths for the \citet{Diamond2001} dataset at Johnstown Castle. The units of $\alpha$ and $K_{sat}$ are m$^{-1}$ and m/day, respectively. All other VG parameters are dimensionless.}
\label{tab:soil_data_JC2}

\begin{tabular}{cccccccccccccc}
\hline
\hline
\multicolumn{2}{c}{Depth (m)} &
\multicolumn{2}{c}{$\theta_s$} &
\multicolumn{2}{c}{$\theta_r$} &
\multicolumn{2}{c}{$\log_{10}(\alpha)$} &
\multicolumn{2}{c}{$\log_{10}(n)$} &
\multicolumn{2}{c}{$\log_{10} (K_{sat})$}  & 
\multicolumn{2}{c}{$\eta$}  \\

Min  & Max & $\mu$ & $\sigma$ & $\mu$ & $\sigma$ & $\mu$ & $\sigma$ & $\mu$ & $\sigma$ & $\mu$ & $\sigma$ & $\mu$ & $\sigma$ \\
\hline
\hline

0.00 & 0.25 & 0.394 & 0.006 & 0.085 & 0.005 & -2.004 & 0.075 & 0.142 & 0.012 & 1.156 & 0.081 & -0.676 & 1.118\\
0.25 & 0.60 & 0.394 & 0.006 & 0.083 & 0.005 & -2.013 & 0.074 & 0.144 & 0.012 & 1.184 & 0.081 & -0.623 & 1.098\\
0.60 & 0.90 & 0.409 & 0.008 & 0.082 & 0.006 & -2.294 & 0.075 & 0.168 & 0.013 & 1.247 & 0.126 & -0.010 & 1.343\\
0.90 & 1.20 & 0.393 & 0.007 & 0.071 & 0.006 & -2.015 & 0.074 & 0.155 & 0.011 & 1.366 & 0.096 & -0.447 & 0.983\\
\hline
\hline
\end{tabular}
\end{table}

\end{landscape}

\begin{landscape}

\begin{table}[htb]
\centering
\small

\caption{Soil texture at different depths -- ISMON dataset}
\label{tab:soil_data_ISMON1}
\begin{tabular}{lcccccc}
\hline
\hline
Site & Depth min (m) & Depth max (m) & Soil type & Sand (\%) & Silt (\%) & Clay (\%) \\
\hline
\hline
\multirow{4}{*}{ISMON dataset} & 0.00 & 0.05 & clay & 21.140 & 22.190 & 56.670 \\
 & 0.05 & 0.10 & clay & 24.432 & 20.922 & 54.646 \\
 & 0.10 & 0.15 & clay & 23.150 & 21.680 & 55.170 \\
 & 0.15 & 0.20 & clay & 23.483 & 21.748 & 54.769 \\
 & 0.20 & 0.25 & clay & 20.387 & 20.110 & 59.503 \\
 & 0.25 & 0.30 & clay & 20.890 & 28.480 & 50.620 \\
\hline
\hline
\end{tabular}
\end{table}

\begin{table}[H]
\small
\centering
\caption{Mean ($\mu$) and standard deviation ($\sigma$) of VG parameters at different depths for the ISMON dataset at Johnstown Castle.   Values of $\theta_s$ in brackets are adjusted values,  to account for the presence of gravel drains at depths between $0.3\,\mathrm{m}$ and $0.5\,\mathrm{m}$.  Ditto marks mean the values are carried down from the previous row. The units of $\alpha$ and $K_{sat}$ are m$^{-1}$ and m/day, respectively. All other VG parameters are dimensionless.}
\label{tab:soil_data_ISMON2}

\begin{tabular}{cccccccccccccc}
\hline
\hline
\multicolumn{2}{c}{Depth (m)} &
\multicolumn{2}{c}{$\theta_s$} &
\multicolumn{2}{c}{$\theta_r$} &
\multicolumn{2}{c}{$\log_{10}(\alpha)$} &
\multicolumn{2}{c}{$\log_{10}(n)$} &
\multicolumn{2}{c}{$\log_{10} (K_{sat})$}  & 
\multicolumn{2}{c}{$\eta$}  \\

Min  & Max & $\mu$ & $\sigma$ & $\mu$ & $\sigma$ & $\mu$ & $\sigma$ & $\mu$ & $\sigma$ & $\mu$ & $\sigma$ & $\mu$ & $\sigma$ \\
\hline
\hline
0.00 & 0.05 & 0.479 & 0.024 & 0.128 & 0.020 & -1.934 & 0.180 & 0.097 & 0.024 & 1.201 & 0.235 & -1.592 & 1.680 \\
0.05 & 0.10 & 0.469 & 0.022 & 0.128 & 0.019 & -1.928 & 0.173 & 0.097 & 0.022 & 1.150 & 0.218 & -1.622 & 1.658\\
0.10 & 0.15 & 0.473 & 0.022 & 0.128 & 0.019 & -1.933 & 0.174 & 0.097 & 0.023 & 1.167 & 0.223 & -1.594 & 1.665\\
0.15 & 0.20 & 0.472 & 0.022 & 0.128 & 0.019 & -1.934 & 0.173 & 0.098 & 0.022 & 1.160 & 0.221 & -1.586 & 1.662\\
0.20 & 0.25 & 0.485 & 0.027 & 0.129 & 0.023 & -1.915 & 0.195 & 0.094 & 0.026 & 1.232 & 0.254 & -1.723 & 1.695\\
0.25 & 0.30 & 0.471 & 0.017 & 0.125 & 0.015 & -2.005 & 0.140 & 0.106 & 0.019 & 1.134 & 0.192 & -1.208 & 1.670\\
     &      & (0.280) &      &      &      &       &       &       &       &       &       &       & \\
0.30 & 0.60 & 0.471 & \ditto & \ditto & \ditto & \ditto & \ditto & \ditto & \ditto & \ditto & \ditto & \ditto  & \ditto\\
     &      & (0.350) &      &      &      &       &       &       &       &       &       &       & \\
0.60 & 2.0 & 0.471 & \ditto & \ditto & \ditto & \ditto & \ditto & \ditto & \ditto &\ditto & \ditto & \ditto  &\ditto\\
\hline
\hline

\end{tabular}

\end{table}

\end{landscape}
\newpage

\section{Results}
\label{sec:results}

In this section we present the results of the three different case studies.   These include a comparison of the RE model with observations, and a cross-comparison  between the RE model, the MoSt model, and the observations.

\subsection{Comparison of RE model with observations}
\label{sec_comparison_RE_obs}

\begin{figure}[H]
\centering
\includegraphics[width=0.9\linewidth]{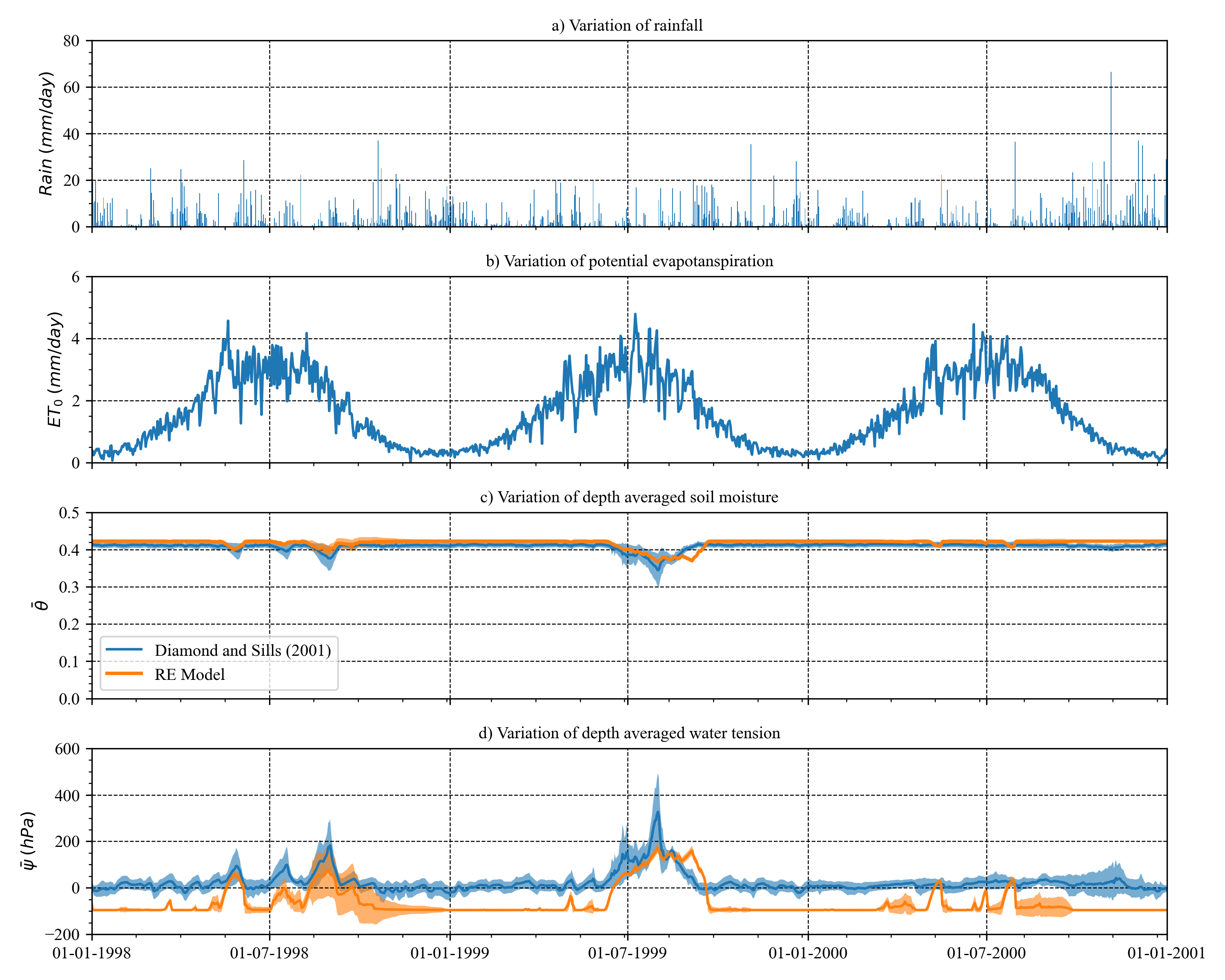}
\caption{Plot of a) daily rainfall $q_r$ (mm/day), b) potential evapotranspiration $(ET_0)$ (mm/day), c) depth-averaged soil moisture $(\bar\theta)$, and d) depth-averaged water tension $(\bar \psi)$ time series for the \citet{Diamond2001} dataset at Clonroche. The line represents the mean ($\mu$) and the shaded zone represents values within one standard deviation ($\sigma$) of the mean.    }
\label{fig:MetClonroche}
\end{figure}

\begin{figure}[H]
\centering
\includegraphics[width=0.9\linewidth]{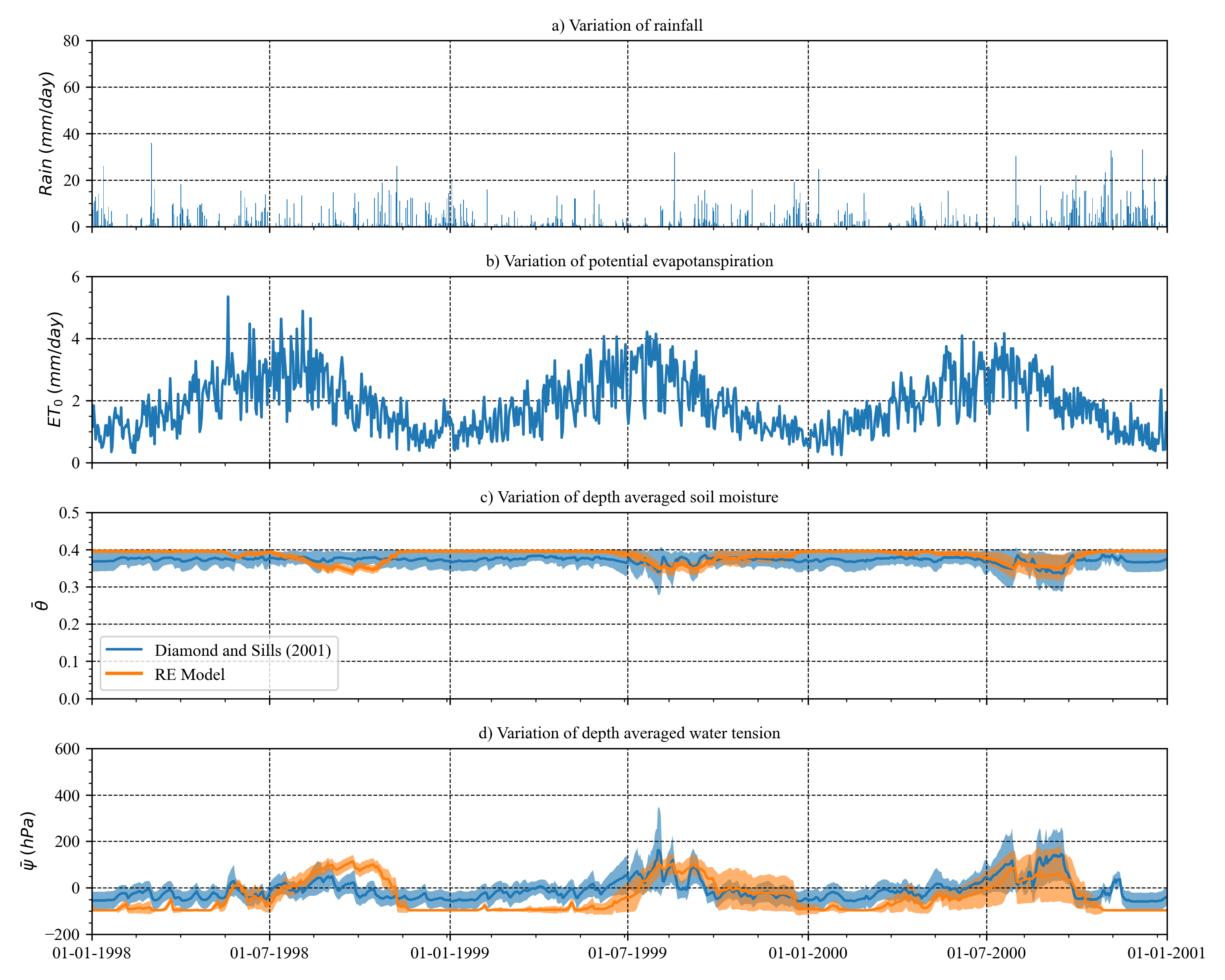}
\caption{Plot of a) daily rainfall $q_r$ (mm/day), b) potential evapotranspiration $(ET_0)$ (mm/day), c) depth-averaged soil moisture $(\bar\theta)$, and d) depth-averaged water tension $(\bar \psi)$ time series for the \citet{Diamond2001} dataset at Johnstown Castle.  }
\label{fig:MetJohnstown}
\end{figure}

\begin{figure}[H]
\centering
\includegraphics[width=0.9\linewidth]{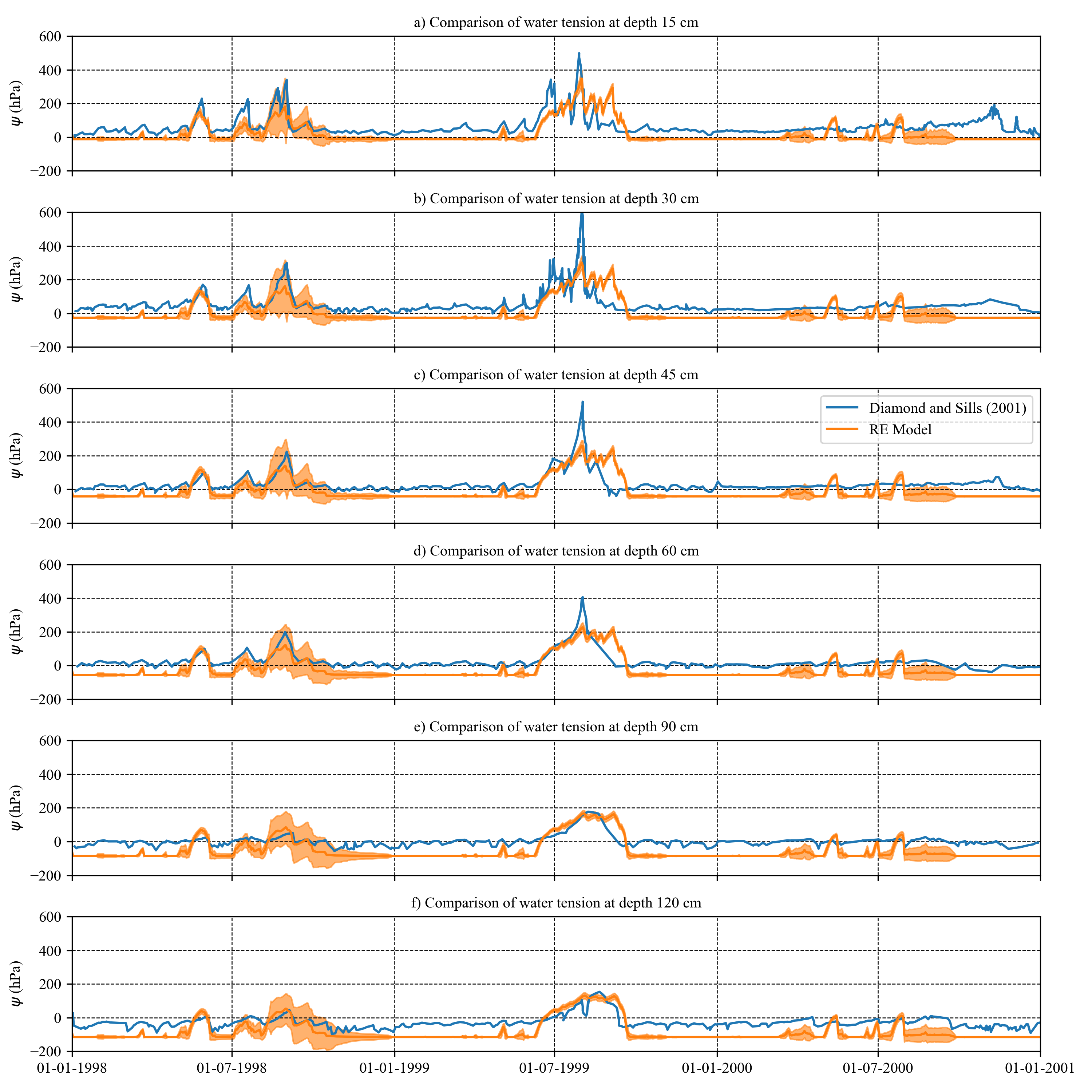}
\caption{ Comparison of predicted water tension (hPa) values at different depths against observed  water tension (hPa) values at Clonroche in the \citet{Diamond2001} dataset}
\label{fig:comparisonTensionClonroche}
\end{figure}

\begin{figure}[H]
\centering
\includegraphics[width=0.9\linewidth]{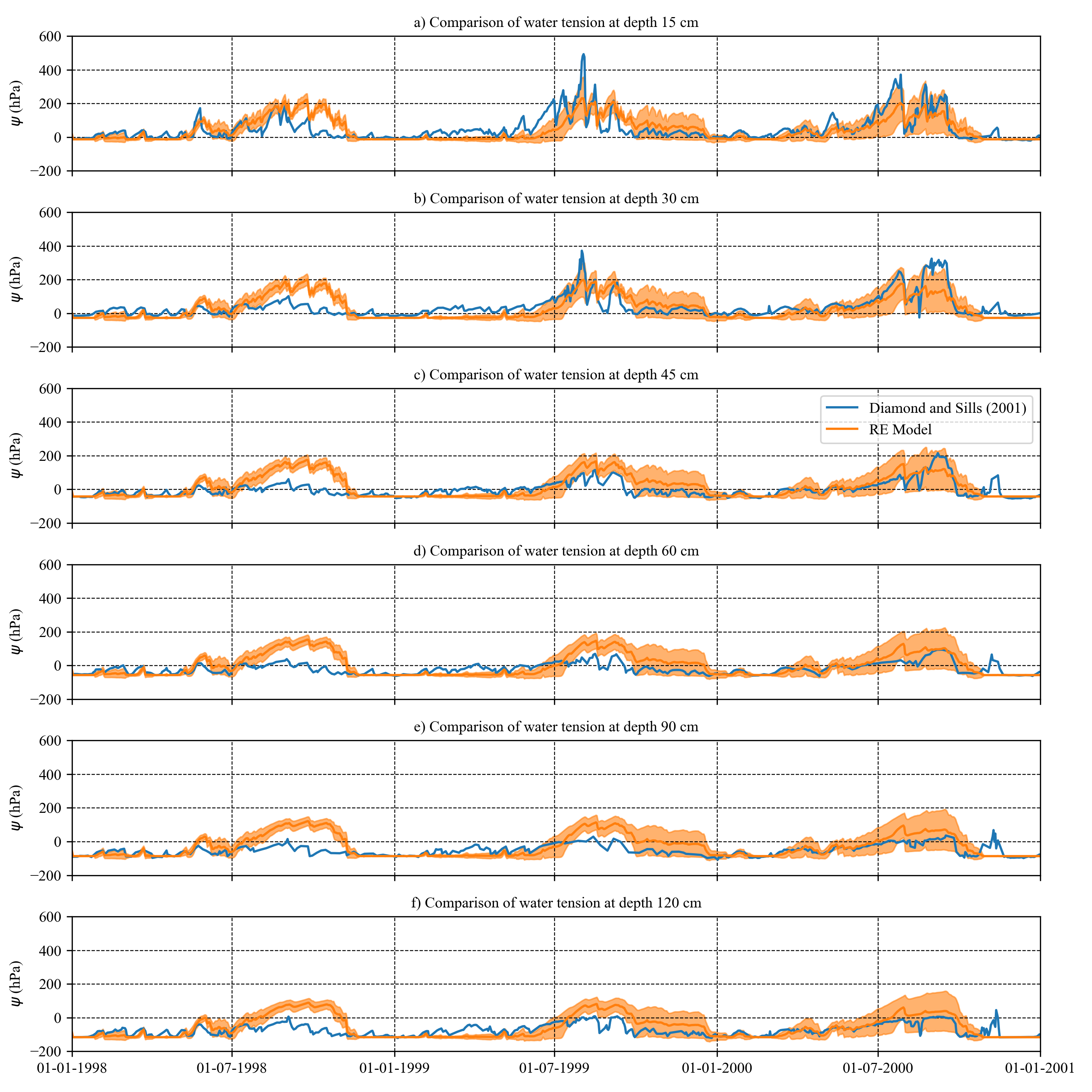}
\caption{Comparison of predicted water tension (hPa) values at different depths against observed  water tension (hPa) values at Johnstown Castle in the \citet{Diamond2001} dataset }
\label{fig:comparisonTensionJohnstown} 
\end{figure}

\paragraph*{Diamond and Sills Dataset:} The meteorological data (daily rainfall $q_r$ and $ET_0$), along with depth-averaged soil moisture $(\bar \theta)$ and water tension $(\bar \psi)$ for Clonroche and Johnstown Castle are presented in summary form in Fig~\ref{fig:MetClonroche} and~\ref{fig:MetJohnstown}, respectively.    Here and throughout the manuscript, we use the format date-month-year for representing time on the horizontal axis.  By plotting the input meteorological variables in the same figure as $\bar\theta$ and $\bar\psi$, the ability of the model to describe the seasonality effects is  highlighted.  In these figures, and throughout the manuscript, we present the mean trajectory of the RE model, along with confidence intervals generated by the Monte Carlo simulation.  Where the observed values lie outside of the confidence intervals, this is explained later in Section \ref{subsec:confidence2}.
We further elaborate on these results by looking not only at depth-averaged values, but also at values of $\theta$ and $\psi$ at precise depths, results of which are shown in Figs~\ref{fig:comparisonTensionClonroche}--\ref{fig:comparisonTensionJohnstown}.

A notable feature is that, although the soils remain close to saturation for large parts of the year, they nevertheless exhibit a clear seasonal cycle. In particular, the observations show modest but consistent drying during the summer months, followed by rewetting during autumn and winter. During the initial stages of this study, the RE model was implemented using the standard Feddes uptake formulation with $f_{2a}=0$. Under these conditions, the Feddes sink term vanished once the anaerobiosis threshold was reached, causing the model to predict near-permanent saturation during extended wet periods. As a result, the observed seasonal drying was not reproduced and substantial discrepancies developed between the model and the observations. This behaviour suggested that residual water losses continue to occur even when soils remain within or close to the anaerobic regime. To account for this effect, a small non-zero value of $f_{2a}=0.01$ was introduced, providing a simple representation of residual water losses from near-saturated soils. 
This modification allowed the model to reproduce the observed seasonal variability and resulted in the improved agreement shown throughout the present section. A systematic sensitivity analysis of the model response to $f_{2a}$ is presented in Appendix~A.

%

Summarizing, the modified RE model reproduces the main seasonal patterns observed at both sites. The model captures the summer drying periods and subsequent winter rewetting. Remaining discrepancies are most evident in the pressure-head variable during prolonged near-saturated conditions, although the corresponding differences in soil moisture are small because these pressure heads lie close to saturation on the water-retention curve. We  emphasize finally that  these results have been produced using a zero-flow bottom boundary condition, as the lower boundary is seen to remain close to saturation during the simulation period.


\begin{figure}[H]
\centering
\includegraphics[width=0.9\linewidth]{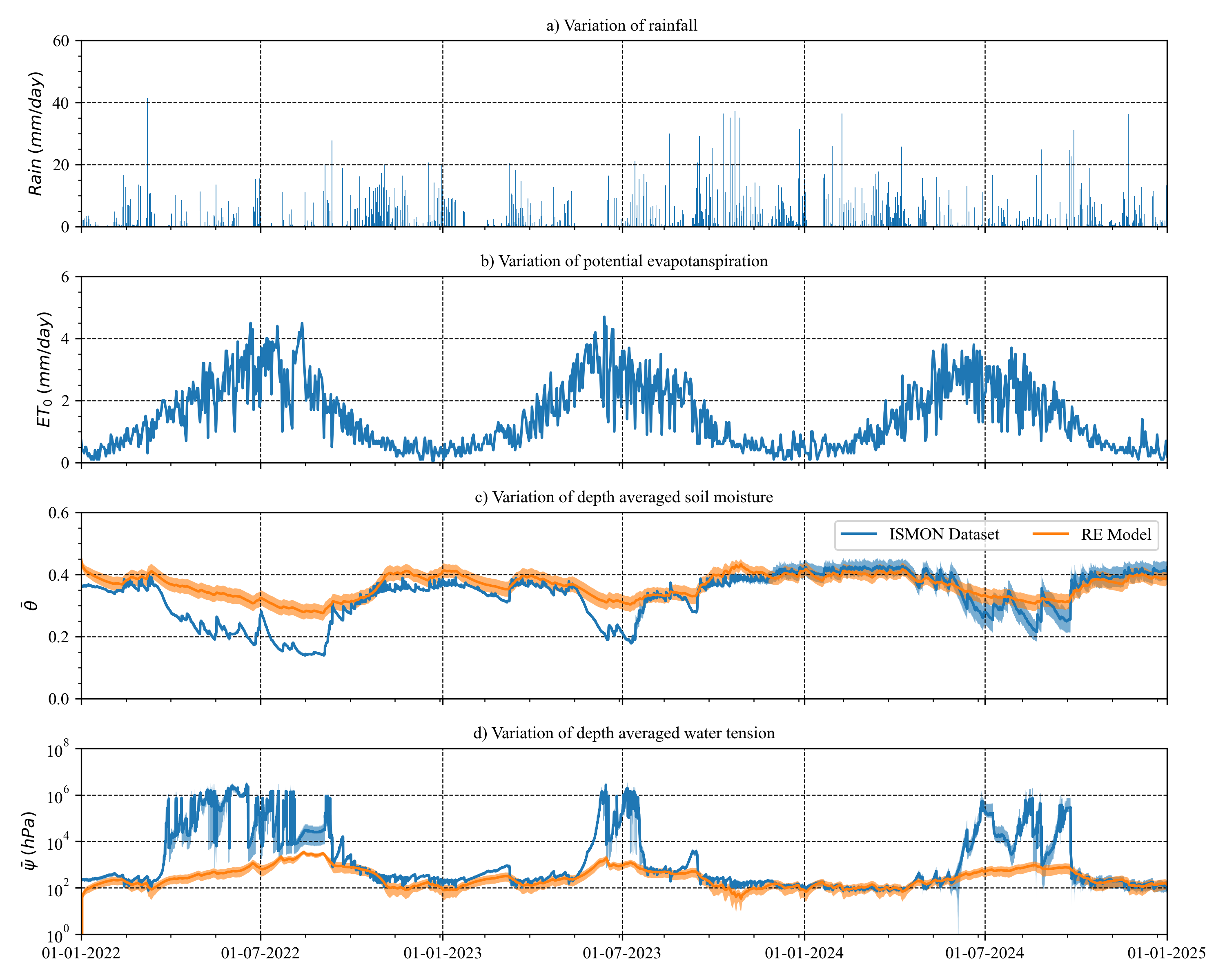}
\caption{Plot of a) daily rainfall $q_r$ (mm/day), b) potential evapotranspiration $(ET_0)$ (mm/day), c) depth-averaged soil moisture $(\bar\theta)$, and d) depth-averaged water tension $(\bar \psi)$ time series for the ISMON dataset at Johnstown Castle.   }
\label{fig:MetISMON}
\end{figure}

\paragraph*{ISMON dataset:}  The meteorological data (daily rainfall and $ET_0$), along with depth-averaged soil moisture $(\bar \theta)$ and water tension $(\bar \psi)$ for Johnstown Castle (ISMON dataset) are presented in summary form in Fig~\ref{fig:MetISMON}.  For the present purposes, the main focus is the $\bar\theta$ variable, since only $\theta$ is directly measured.  

An important distinction between the ISMON dataset and the two earlier case studies is that the soil profile does not remain very close to saturation throughout the year. Although near-saturated conditions occur during the winter months, substantial drying is observed during the summer period, extending throughout much of the soil column. Consequently, the model behaviour is less strongly controlled by the treatment of anaerobic conditions within the Feddes uptake formulation. Unlike the previous case studies, where the standard choice $f_{2a}=0$ led to near-permanent saturation and a failure to reproduce the observed seasonal drying, the dominant discrepancies in the ISMON dataset arise during the dry summer periods. These discrepancies are discussed below and are primarily associated with the representation of residual soil moisture rather than the treatment of water losses under anaerobic conditions.
This contrast with the previous case studies supports the hypothesis that the treatment of anaerobic conditions becomes most important when soils remain close to saturation for extended periods.

The RE model in Fig~\ref{fig:MetISMON} captures the seasonality effect in the observations -- significant dryout in the summers and saturation or near-saturation in the winters.  Very close agreement is not possible in the summer periods, which we comment on below.  Also, very close agreement in the $\bar\psi$ variable is not possible, however, this is not problematic since only $\theta$ is of interest in the present study.  Furthermore, this discrepancy can be understood as a by-product of the postprocessing, which is 
used to convert observations of $\theta$ to estimates of $\psi$.  The postprocessing is done by passing the measured value of $\theta$ through the inverse VG function.  As this function possesses strong gradients, small discrepancies in $\theta$ are magnified by the inverse VG function, to produce much larger discrepancies in $\psi$.
The remaining discrepancies during the dryer summer periods therefore point towards limitations in the representation of residual soil moisture.

We present a more detailed study in Fig~\ref{fig:comparisonISMONJohnstown} where we look at the soil moisture at different depths.  Again, the RE model predicts the trends in the data at all depths.  Here, we use a free drainage bottom boundary condition at $L=2\,\mathrm{m}$, since the lower boundary does not remain close to saturation for the entire simulation period.  
Very close agreement is not possible: in particular, the observations indicate dryout in the summer periods, with $\theta$ reaching as low as $0.1$ in August 2022, at depths ranging from $5\,\mathrm{cm}$ to $20\,\mathrm{cm}$.  In contrast, the RE model only attains values of $\theta \approx 0.2$ in the same periods, at the same depths.  

It is clear from these results that the model experiences $\theta\approx 0.2$ as an effective lower bound.  To mitigate the effect of this effective lower bound, we have tried systematically reducing the parameter value $\theta_r$ down to $0.01$ for each soil layer. By re-running the simulation, there is much better qualitative and quantitative agreement between the model and the observations.  To maintain the readability of the  manuscript, results of this brief parameter study are presented in Appendix~\ref{sec:app:varythetar}. 

Although the adjusted value of $\theta_r$ lies outside the uncertainty range predicted by ROSETTA~3 (e.g. Table~\ref{tab:soil_data_ISMON2}), there are good reasons to expect discrepancies between the pedotransfer-function estimates and the effective hydraulic properties of the monitored site. As documented in Section~\ref{sec:MM}, the TDR sensors producing the observations were placed in pits which were dug out and subsequently refilled. Gravel drainage layers were also present at the site in the ISMON dataset, further altering the hydraulic properties of the soil profile. Consequently, the hydraulic properties inferred from soil texture alone may not fully represent the effective hydraulic behaviour governing the observations. This provides a plausible explanation for why improved agreement is obtained when $\theta_r$ is reduced relative to the value predicted by ROSETTA~3.
Unlike the Feddes-related discrepancies identified in the Diamond and Sills datasets, the discrepancies in the ISMON dataset appear to be associated primarily with uncertainty in site-specific hydraulic parameters.

Similar adjustment was made to $\theta_s$ to account for presence of drainage layers at depths ranging from $30\,\mathrm{cm}$ to $50\,\mathrm{cm}$. As shown in Fig~\ref{fig:comparisonISMONJohnstown}, the maximum value of the observed $\theta$ did not exceed $0.30$ and $0.35$ at depths $30\,\mathrm{cm}$ and $50\,\mathrm{cm}$ respectively. These reduced maximum $\theta$-values are consistent with the presence of gravel drainage layers, which effectively decrease the water-holding capacity of the soil profile. Consequently, the values of $\theta_s$ for the affected layers were reduced accordingly. The modified values of $\theta_s$ are reported in Table~\ref{tab:soil_data_ISMON2}.


\subsection{Confidence Intervals}
\label{subsec:confidence2}

\begin{figure}[H]
\centering
\includegraphics[width=0.9\linewidth]{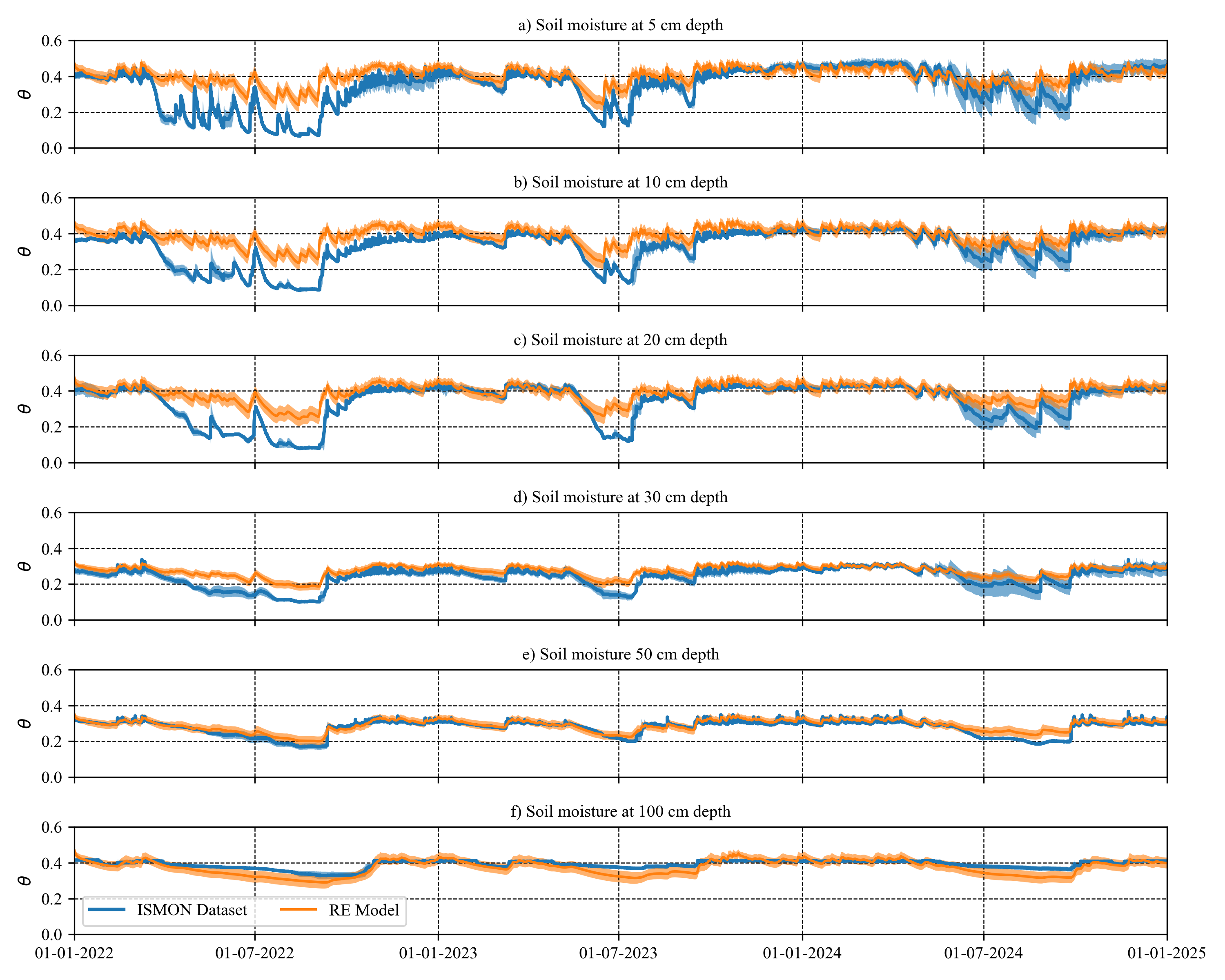}
\caption{Comparison of predicted soil moisture, $\theta$ values at different depths against observed soil moisture, $\theta$ at Johnstown Castle for the ISMON dataset.}
\label{fig:comparisonISMONJohnstown} 
\end{figure}

The ISMON results discussed above provide one example of how site-specific hydraulic properties may fall outside the uncertainty represented by ROSETTA 3.
Reasoning along the same lines, the confidence intervals shown throughout the present article (in particular, Figs~\ref{fig:comparisonTensionClonroche}--\ref{fig:comparisonISMONJohnstown}) are derived from assumptions on the soil hydraulic properties in ROSETTA 3. We note that ROSETTA 3 is an empirical pedotransfer function model and is thus limited in its prediction accuracy. The uncertainties in VG parameters are primarily from soil textural information~\citep{ZHANG201739} and do not account for site-specific disturbances. The presence of artificial drains or human activities leads to significant changes in soil properties. Site-specific disturbances such as presence of drains or soil excavation, and structural heterogeneity are not explicitly accounted for in ROSETTA 3. Hence, any variation in soil hydraulic properties due to local effects is not accounted for by these confidence intervals.   This explains why the observations are occasionally out of the bounds implied by the confidence intervals -- particularly in the dry summers.


\subsection{Cross-comparison: RE model, observations, and MoSt model}

We present here a cross-comparison between the RE model, the observations, and the MoSt model.  Because the MoSt model only gives values for the water storage values at $10\,\mathrm{cm}$ ($\Theta_{10}$) and $100\,\mathrm{cm}$ ($\Theta _{100}$), our comparison is based on these quantities.   The MoSt model also assumes a free-drainage condition at the bottom of the soil column.  As this assumption is only appropriate for the ISMON-Johnstown Castle dataset, we restrict the comparison to the ISMON-Johnstown Castle dataset (for accurate modelling of the waterlogging observed in the Diamaond and Sills dataset, a no-flow boundary condition was required at the bottom of the soil column, which is not possible in the MoSt model).

\begin{figure}[]
\centering
\includegraphics[width=0.9\linewidth]{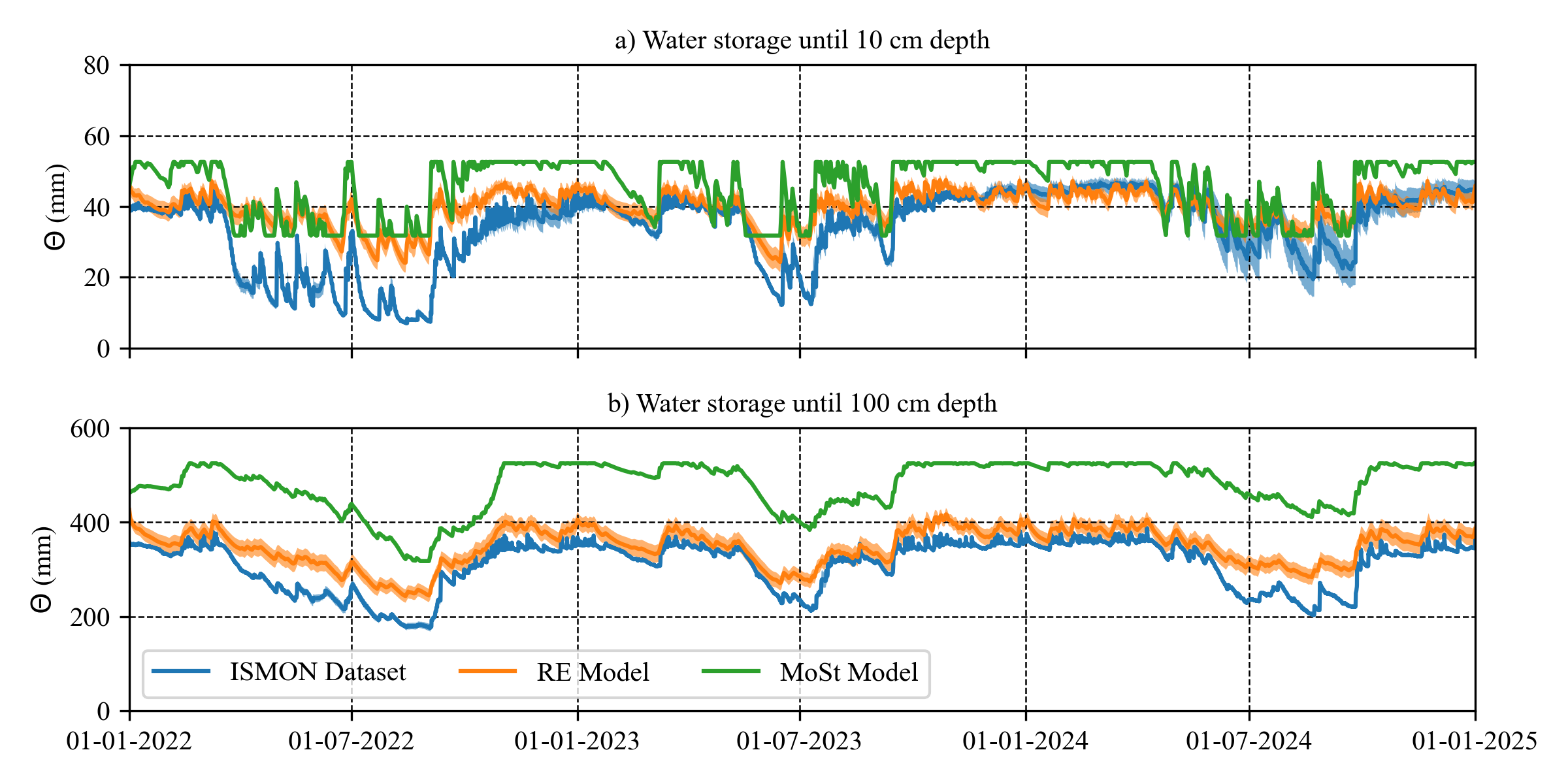} 
\caption{Comparison of predicted water storage values until 10 cm depth and 100 cm depth against observations at Johnstown Castle in the ISMON dataset.} 
\label{fig:Comparison_water_storageISMON} 
\end{figure}

Results are shown in Fig~\ref{fig:Comparison_water_storageISMON}.   They demonstrate 
a pronounced seasonal drying throughout the soil profile.  Model performance is not sensitive to the treatment of anaerobic conditions.
Instead, the dominant discrepancies are associated with uncertainty in the hydraulic properties of the site.  The results also show a `shift' or a systematic bias in the predictions of the MoSt model compared both to the data and the RE model.

\begin{figure} []
\centering
\includegraphics[width=0.9\linewidth]{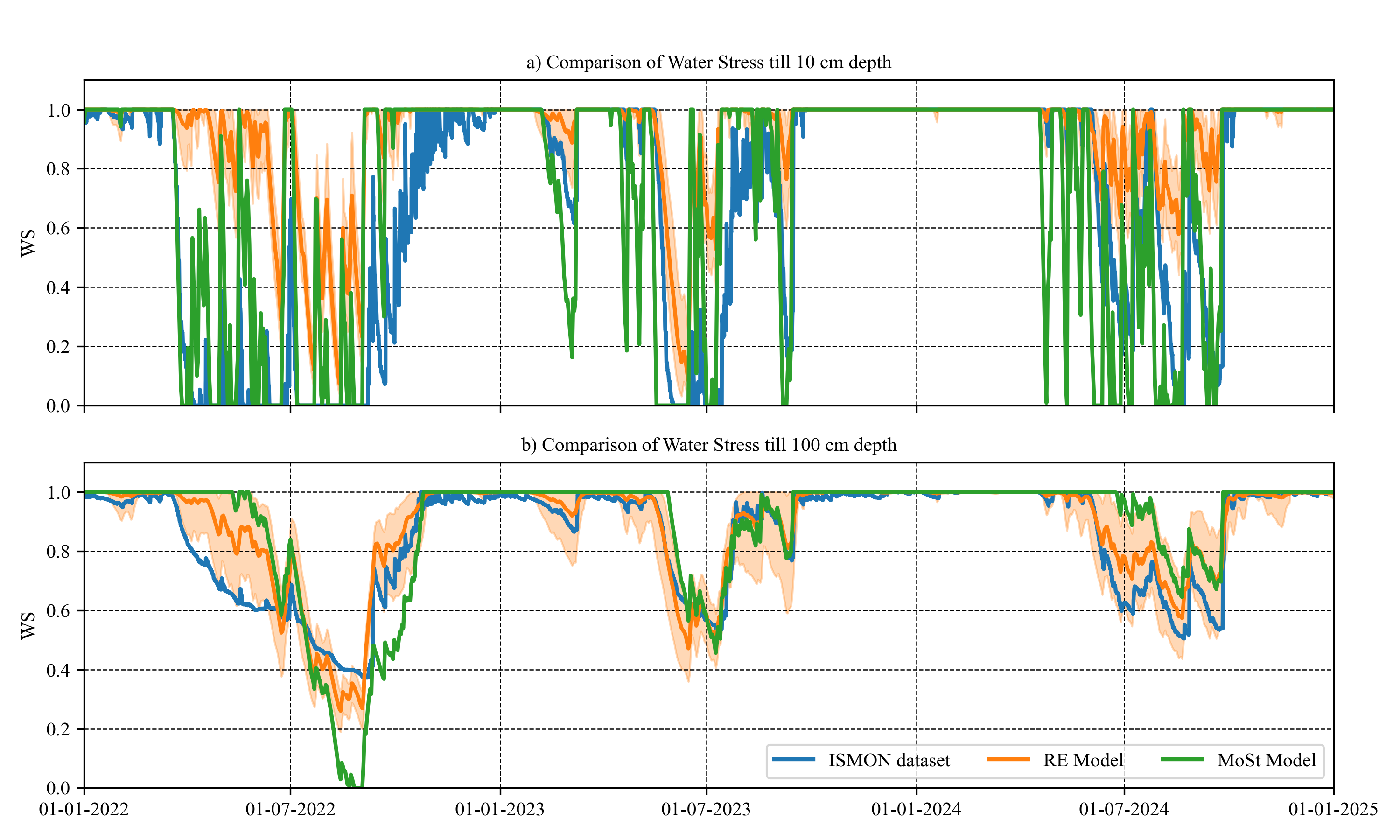} 
\caption{Water stress (RE model and MoSt model) at different depths for Johnstown Castle in the ISMON dataset.} 
\label{fig:waterstress_comparison} 
\end{figure}

In practice, the bias does not affect the grass-growth predictions, as these are based on the water stress function. The water stress (WS) at arbitrary depth $z$ is defined as
\begin{equation}
\ws(z,t)=\min\left(1,\left|\frac{\theta(z,t)-\wlp}{\fc-\wlp}\right|\right),
\label{eq:ws}
\end{equation}
where $\wlp$ is the wilting point, and $\fc$ is the field capacity (all dimensionless quantities). These can be computed using ROSETTA-3, and are layer-dependent, similar to the VG parameters in Tables~\ref{tab:soil_data_CR2}, \ref{tab:soil_data_JC2}, and~\ref{tab:soil_data_ISMON2}. Here, FC and WP were estimated from the soil water retention curve at pressure heads $h=-3.3$ m and $h=-150$ m, respectively.
The water stress is between 0 and 1 and can be interpreted as a normalised measure of plant-available water.


Based on Equation~\eqref{eq:ws}, we have computed $\ws(z,t)$ and hence, $\ws_{10}(t)$ and $\ws_{100}(t)$ for the RE model and the ISMON dataset.
These involve the depth-averaged values of $\theta(\cdot,t)$ at $10\,\mathrm{cm}$ and $100\,\mathrm{cm}$ respectively. The $WS$-value for the MoSt model was also obtained, for a depth of $100\,\mathrm{cm}$. We compare all three time series in Fig~\ref{fig:waterstress_comparison}. The sharp discrepancy between the RE model and the MoSt model previously observed in Fig~\ref{fig:Comparison_water_storageISMON} is no longer apparent -- the bias is removed by the shift in $\theta$-values produced by Equation~\eqref{eq:ws}.
The figure shows  broad agreement (both qualitative and quantitative) between the MoSt model and the RE model.  
The performance of the MoSt and RE models was evaluated by calculating the root mean square error (RMSE) between the model predictions and the $WS$ estimates derived from the ISMON-Johnstown Castle dataset. Comparisons were performed for the averaged $\ws(z,t)$ up to depths of 10 cm  and 100 cm soil profile. For the 10 cm layer, the MoSt model showed better agreement with the observations, with an RMSE of $0.264$, compared with $0.371$ for the RE model. In comparison, for the averaged $\ws(z,t)$ up to 100 cm depth, the RE model outperformed the MoSt model, resulting in a lower RMSE $(0.075)$ than the MoSt model $(0.134)$. These results indicate that the RE model provides a more accurate representation of integrated root-zone water stress, whereas the MoSt model more effectively captures the variability in near-surface water stress.
A main difference between the MoSt model and the RE model is that the plant stress does not go to zero in summer 2022, which has implications for grass-growth predictions.

\section{Discussion and Conclusions}
\label{sec:disc}

Summarizing, we have developed an RE model for predicting soil moisture for three different case studies in Ireland, two of which demonstrate near-year-long waterlogging. The model predicts well the seasonal trends in depth-averaged soil moisture $(\bar{\theta})$ and depth-averaged water tension $(\bar{\psi})$. The model is able to estimate $\bar{\theta}$ more accurately than $\bar{\psi}$ and $\bar h$ across all three case studies.  This is a useful result, as the key variable of interest for the present purposes (in particular, as an input to grass-growth models) is precisely $\bar{\theta}$.

A key finding of this study is that the largest discrepancies between the RE and the observations occur during prolonged periods of near-saturation. We suggest that these discrepancies may arise from the treatment of anaerobic conditions within the standard Feddes uptake formulation. In the conventional formulation, plant water uptake is completely suppressed once the pressure head exceeds the anaerobiosis threshold. While this assumption is intended to represent oxygen limitation of plant roots, it neglects continued water losses that may occur from near-saturated soils. As a result, the standard formulation systematically over-predicts soil moisture under waterlogged conditions. Introducing a small non-zero value of $f_{2a}$ substantially improves agreement with observations across the different case studies, supporting the hypothesis that the standard treatment of anaerobic conditions within the Feddes uptake formulation is inadequate for persistently wet grassland systems.

A second important finding concerns the role of soil hydraulic properties in controlling
model performance. While the discrepancies observed when comparing RE with the Diamond and Sills datasets
were primarily associated with the treatment of anaerobic conditions, the dominant source
of error in the ISMON dataset arose during the dry summer periods and was linked to the
representation of residual soil moisture. In particular, reducing the parameter $\theta_r$
produced substantially improved agreement with observations, suggesting that the effective
hydraulic properties of the monitored soil profile differ from those predicted by the
ROSETTA~3 pedotransfer function. This is perhaps unsurprising given the presence of
gravel drains and the disturbance associated with sensor installation. More generally, these
results highlight the sensitivity of the RE to hydraulic parameter
estimation and demonstrate that uncertainty in soil properties can be as important as the assumptions used to represent water losses under anaerobic conditions.
Taken together, the Diamond and Sills and ISMON case studies demonstrate that the dominant source of model error depends on the prevailing hydrological regime, with anaerobic conditions controlling performance in persistently wet soils and hydraulic parameter uncertainty becoming more important in seasonally drying soils.

The comparison with the operational MoSt model  has implications for grass-growth forecasting.  The most pertinent comparison is with respect to the water stress.  The main finding is that the $f_{2a}$-corrected RE model is consistent with the observations across all three study sites, 
whereas the MoSt model is consistent with the observations only in case of non-waterlogged soils (ISMON case study).  Both the MoSt model and the RE model may be improved  by further calibrating input parameters.  However, one of the main aims of the paper is the development of a model with the ability to make \textit{a priori} predictions of soil moisture, with minimal need for calibration of parameters.  For that reason, `fine-tuning' of the model parameters (especially VG parameters) is not looked in the present article, although it may be of interest for future work.

While the introduction of a non-zero value of $f_{2a}$ improves model performance, it should be regarded as a pragmatic approximation rather than a physically rigorous representation of water losses in waterlogged soils.
The performance of the RE model could be further improved through better characterization of soil hydraulic properties and improved representation of hydrological processes such as evaporation, plant water uptake, etc. The partition of $ET_0$ into evaporation and transpiration components may also improve model performance, although this was partially mitigated through modification of the Feddes function.  Furthermore, the discrepancies between the RE model outputs and observations during drying periods may be partly due to the use of pedotransfer functions to estimate VG parameters. Pedotransfer functions such as ROSETTA~3 do not fully represent soil structure, preferential flow pathways, macroporosity, biological activity, or shrink-swell behaviour, all of which may substantially influence field-scale hydraulic behaviour.
While the Monte Carlo framework accounted for uncertainty in VG parameters, temporal variability in soil hydraulic properties was not considered.  In addition, the use of a 1D RE model neglects lateral flow, which may be important in poorly drained or artificially drained soils.
All of these effects could be explored in a future possible study. Nevertheless, for the present aim of developing a relatively simple model capable of being deployed to make real-time soil moisture predictions of use both to researchers and farmers, the present 1D approach is sufficient.
	
These findings highlight the value of physically-based soil-moisture models for bridging the gap between understanding hydrological processes and managing grasslands. 
The RE model delivers new insights into the performance of the MoSt model, and can be used to inform parameter selection within same.
Beyond their relevance to Irish grasslands, the results identify an important limitation of conventional Feddes-type uptake formulations in waterlogged soils and demonstrate a simple practical approach for improving predictions under prolonged wet conditions. Improved representation of soil-moisture dynamics in such environments can support more reliable grass-growth forecasting, more informed management decisions, and ultimately more productive and environmentally sustainable grassland systems.
    

\subsection*{Data availability statement}

The historical meteorological data for all sites investigated are publicly available (\url{https://www.met.ie/climate/available-data/historical-data}).
The dataset by Diamond and Sills is publicly available~\citep{Diamond2001}.  Digitized values from this dataset are available for download from the Github repository of Saurabh Kumar
(\url{https://github.com/ksaurabh95/1D_RE_Model_MonteCarlo.git}).  The code for the RE model is available for download at the same repository.

\section*{Declaration of Competing Interest}

The authors declare that they have no known competing financial interests or personal relationships that could have appeared to influence the work reported in this paper.

\subsection*{Acknowledgments}
This publication has emanated from research conducted with the financial support of Research Ireland and the Department of Agriculture, Food and Marine on behalf of the Government of Ireland under Grant Number [21/RC/10303P2] -- VistaMilk.
We are thankful to ISMON for providing the soil moisture dataset for Johnstown Castle Research station. Furthermore, we are thankful to Met \'Eireann for the various meteorological datasets used in this study. 
We would also like to thank Tamara Hochstrasser for helpful discussions.

\clearpage

\newpage
\appendix

\section{Sensitivity of the RE model to the anaerobic uptake parameter $f_{2a}$}

\label{sec:app:varyf_2a}

A central finding of this study is that the treatment of anaerobic conditions strongly
influences model performance in persistently wet grassland systems. In the standard
Feddes formulation~\eqref{eq:f1def}, the parameter $f_{2a}$ is set to zero, implying
that the sink term vanishes once the pressure head exceeds the anaerobiosis threshold
$h_a$. As discussed in the main text, this led to systematic over-prediction of soil
moisture in the Diamond and Sills datasets.

To investigate the sensitivity of the model to this assumption, a parameter study was
performed using the Johnstown Castle case study from the Diamond and Sills dataset.
Values of $f_{2a}$ between 0 and 0.5 were considered while all other model parameters
were held fixed. The resulting model predictions are shown in Figure~\ref{fig:Comparison_soil_moistureISMON_revised_thetar}.

When $f_{2a}=0$, the model predicts near-permanent saturation and fails to reproduce
the observed seasonal drying. 
As $f_{2a}$ is increased above zero, the model allows limited water extraction under
near-saturated conditions, leading to improved agreement with the observations up to
about $f_{2a}=0.01$. For larger values of $f_{2a}$, the model tends to over-dry the soil,
and the agreement with observations deteriorates.
These results support the
hypothesis that complete suppression of the sink term under anaerobic conditions is
inappropriate for persistently wet grassland systems.

Based on this analysis, a value of $f_{2a}=0.01$ was adopted in the main simulations.
The purpose of the parameter study is not to identify a unique optimal value of $f_{2a}$,
but rather to demonstrate the importance of accounting for residual water losses under
near-saturated conditions.

\begin{figure} [htb]
\centering
\includegraphics[width=0.9\linewidth]{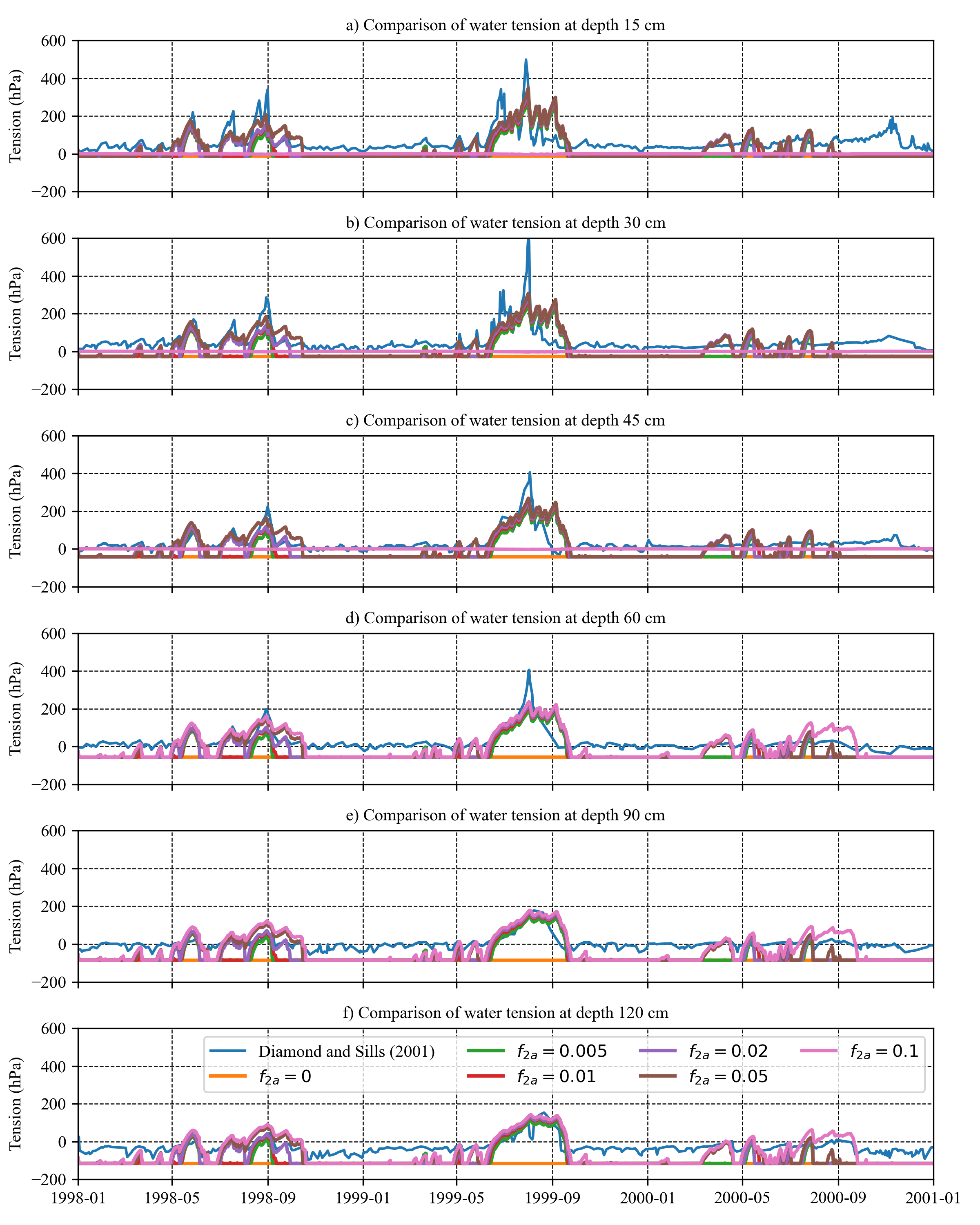} 
\caption{Comparison of RE predicted water tension values for all depths with varying $f_{2a}$, against observations at Clonroche Castle in the Diamond and Sills (2001) dataset.} 
\label{fig:Comparison_soil_moistureISMON_revised_thetar} 
\end{figure}

\begin{figure} [htb]
\centering
\includegraphics[width=0.9\linewidth]{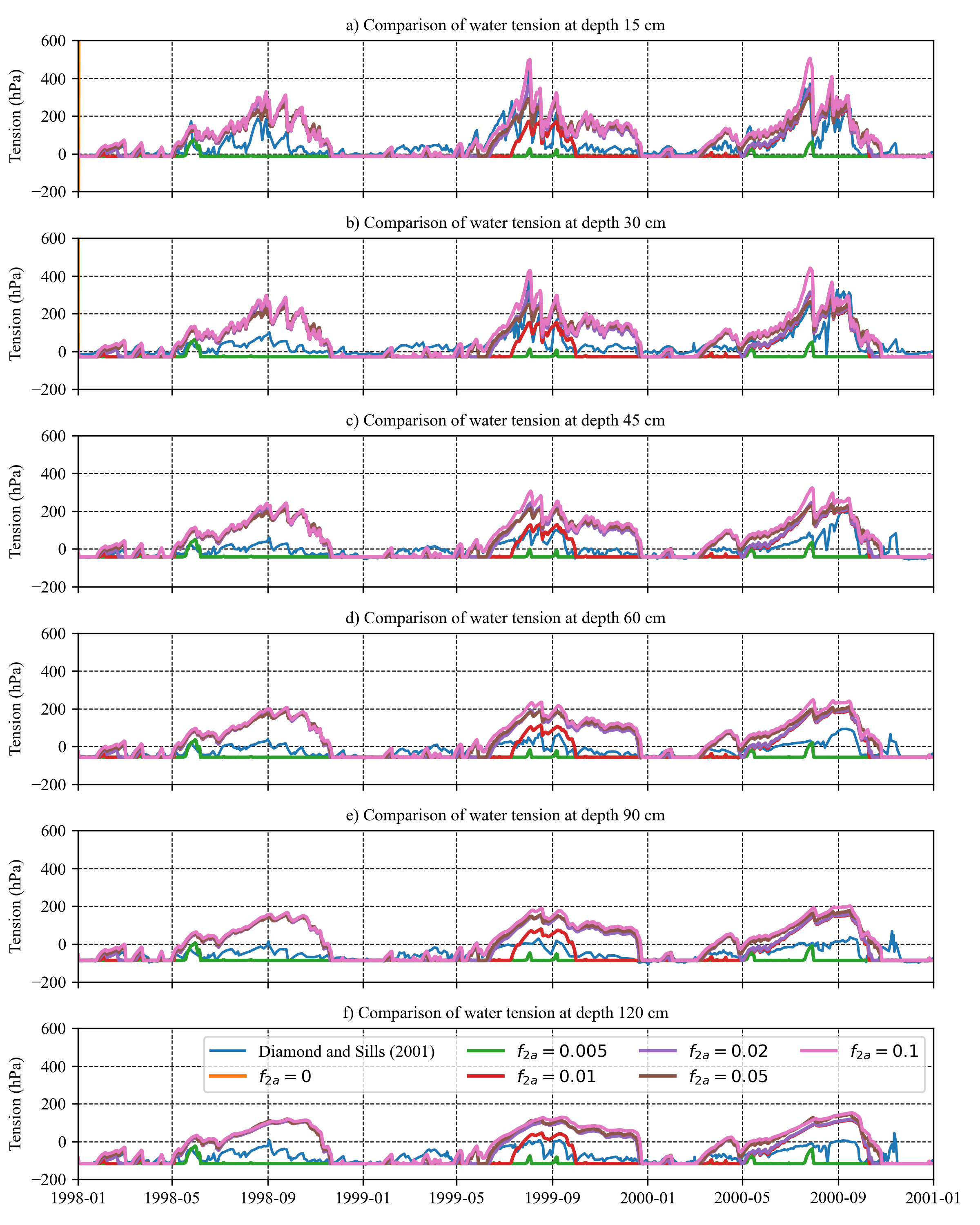} 
\caption{Comparison of RE predicted water tension values for all depths with varying $f_{2a}$, against observations at Johnstown Castle in the Diamond and Sills (2001) dataset.} 
\label{fig:Comparison_soil_moistureISMON_revised_thetar} 
\end{figure}

\begin{figure} [htb]
\centering 
\includegraphics[width=0.9\linewidth]{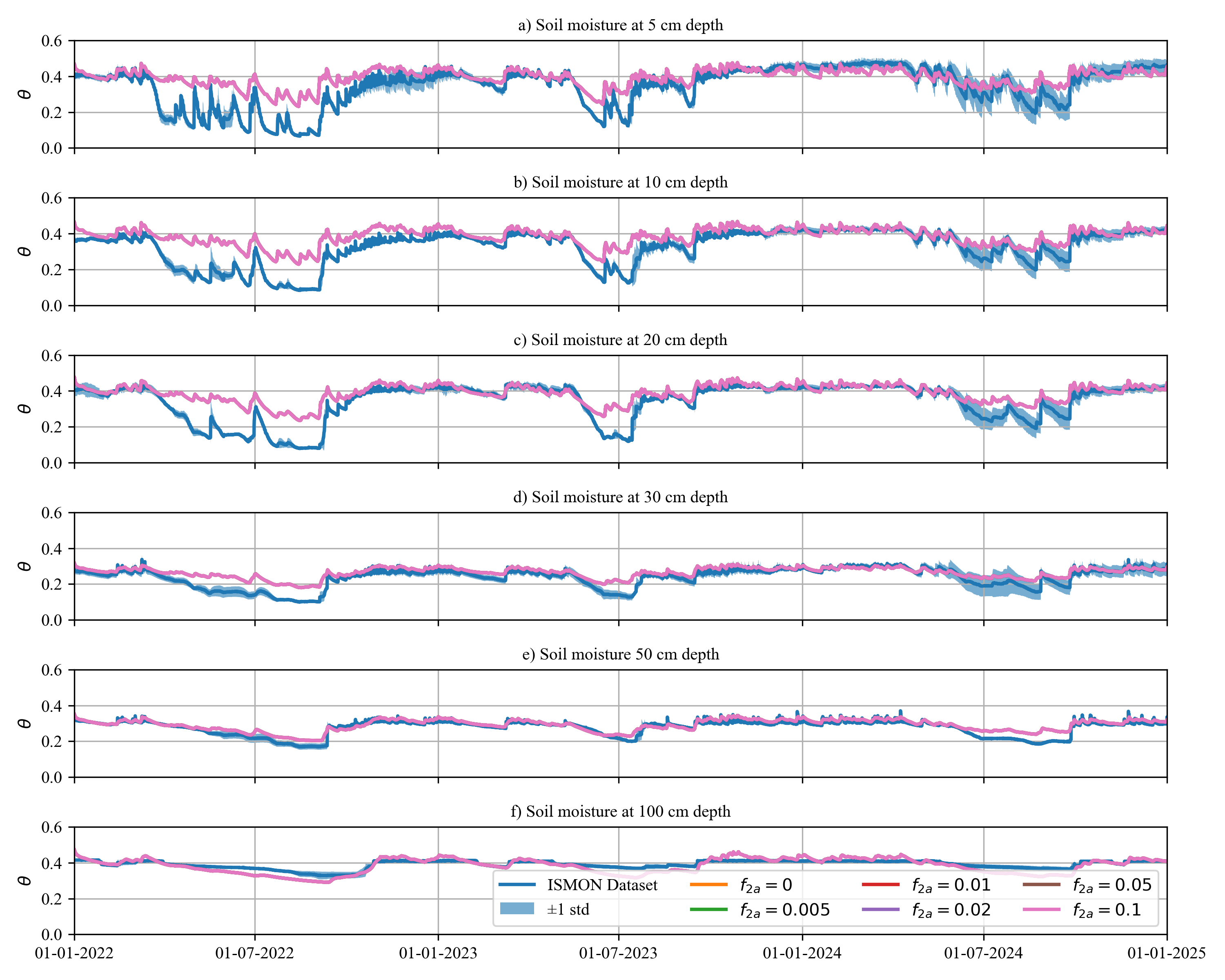} 
\caption{Comparison of RE predicted water tension values for all depths with varying $f_{2a}$, against observations at Johnstown Castle in the ISMON dataset.} 
\label{fig:Comparison_soil_moistureISMON_revised_thetar} 
\end{figure}

\clearpage
\newpage

\section{Parameter study showing the effect of varying $\theta_r$ (ISMON dataset)}
\label{sec:app:varythetar}

In Section~\ref{sec_comparison_RE_obs}, the effect of varying $\theta_r$ in the RE model was discussed in the context of the ISMON dataset.  Here we re-run the RE model for Johnstown Castle with parameters and meteorological data as per the ISMON dataset. We looked at the effect of reducing the residual moisture level to $\theta_r=0.01$ throughout the soil column.  The results  of this investigation are shown below in Fig~\ref{fig:Comparison_soil_moistureISMON_revised_thetar2}.
The results demonstrate that the RE is sensitive to the choice
of the residual soil moisture parameter, $\theta_r$. Reducing $\theta_r$ lowers the effective
minimum soil moisture attainable by the model and allows the stronger summer dryout
observed in the ISMON dataset to be reproduced more accurately. This improvement
supports the interpretation presented in Section~\ref{sec_comparison_RE_obs} that the dominant discrepancies in
the ISMON dataset are associated with uncertainty in the hydraulic properties of the site,
rather than the treatment of anaerobic conditions within the Feddes uptake formulation.

\begin{figure}[htb]
\centering
\includegraphics[width=0.9\linewidth]{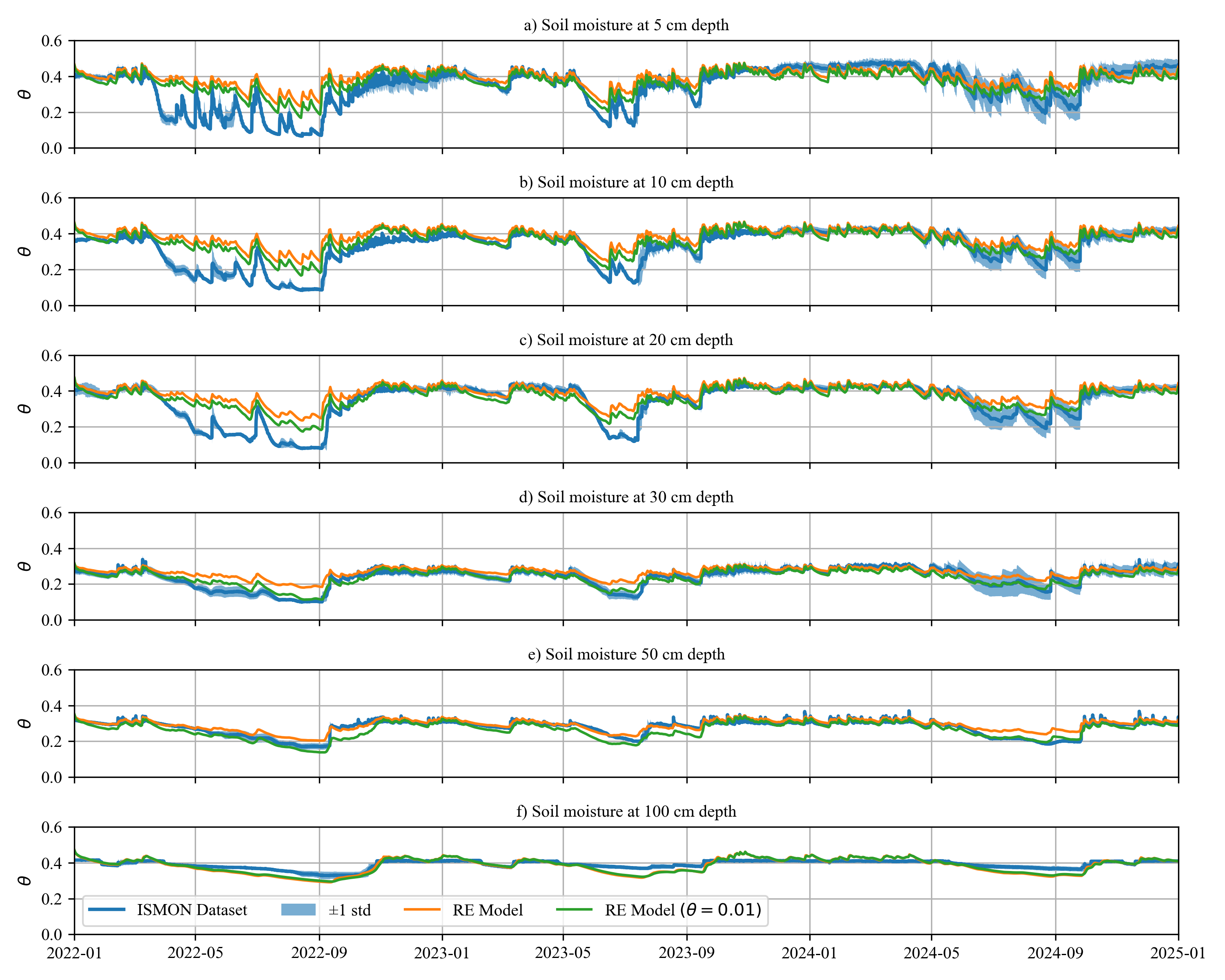} 
\caption{Comparison of RE predicted soil moisture values for all depths with reduced $\theta_r = 0.01$, against observations at Johnstown Castle in the ISMON dataset.} 
\label{fig:Comparison_soil_moistureISMON_revised_thetar2} 
\end{figure}

%
%

\bibliographystyle{apalike}
\bibliography{apssamp}

\end{document}